\documentclass[10pt,journal]{IEEEtran}

\usepackage{amsthm}
\usepackage{times}
\usepackage{multirow}
\usepackage{color,amsmath,amssymb,url,listings,subfig,graphicx,setspace,comment}
\usepackage{textcomp}
\usepackage[ruled,vlined]{algorithm2e}
\usepackage{cite} 

\newtheorem{theorem}{Theorem}
\newtheorem{lemma}{Lemma}
\newtheorem{proposition}{Proposition}

\newtheorem{definition}{Definition}
\makeatletter
\newtheorem*{rep@theorem}{\rep@title} \newcommand{\newreptheorem}[2]{%
\newenvironment{rep#1}[1]{%
\def\rep@title{\bf #2 \ref{##1} }%
\begin{rep@theorem} }%
{\end{rep@theorem} } }
\makeatother
\newreptheorem{theorem}{Theorem}
\newreptheorem{lemma}{Lemma}

\begin{document}

\title{TrueTop: A Sybil-Resilient System for User Influence Measurement on Twitter
\thanks{J. Zhang, J. Sun, and Y. Zhang are with the Department of
 Electrical, Computer, and Energy Engineering, Arizona State University, Tempe, AZ 85287 (e-mail:
{jxzhang, jcsun, yczhang}@asu.edu).}
\thanks{R. Zhang is with the Department of Electrical Engineering, University of
Hawaii, Honolulu, HI 96822 (e-mail: ruizhang@hawaii.edu).}
\thanks{C. Zhang is with the Key Laboratory of Electromagnetic Space Information, Chinese Academy of Sciences, and the School of Information Science and Technology, University of Science and Technology of China, Hefei 230026, China (email:chizhang@ustc.edu.cn).}
}

\author{Jinxue~Zhang,~\IEEEmembership{Student~Member,~IEEE,}
        Rui~Zhang,~\IEEEmembership{Member,~IEEE,}
        Jingchao~Sun,~\IEEEmembership{Student~Member,~IEEE,}
        Yanchao~Zhang,~\IEEEmembership{Senior~Member,~IEEE,}
        and~Chi~Zhang,~\IEEEmembership{Member,~IEEE}}

\maketitle


\begin{abstract}
Influential users have great potential for accelerating information dissemination and acquisition on Twitter. How to measure the influence of Twitter users has attracted significant academic and industrial attention. Existing influence measurement techniques are vulnerable to sybil users that are thriving on Twitter. Although sybil defenses for online social networks have been extensively investigated, they commonly assume unique mappings from human-established trust relationships to online social associations and thus do not apply to Twitter where users can freely follow each other. This paper presents TrueTop, the first sybil-resilient system to measure the influence of Twitter users. TrueTop is rooted in two observations from real Twitter datasets. First, although non-sybil users may incautiously follow strangers, they tend to be more careful and selective in retweeting, replying to, and mentioning other users. Second, influential users usually get much more retweets, replies, and mentions than non-influential users. Detailed theoretical studies and synthetic simulations show that TrueTop can generate very accurate influence measurement results with strong resilience to sybil attacks.
\end{abstract}


\section{Introduction}\label{sec:Intro}
\IEEEPARstart{T}{witter} has become a powerful vehicle for large-scale information dissemination. As of May 2014, Twitter has 255 million monthly active users and 500 million daily tweets. This massive base of active users has triggered explosive uses of Twitter in marketing, journalism, public relations, massive information campaigns, entertainment, and during events of worldwide and national significance.

Influential Twitter users have great potential for accelerating information dissemination and acquisition. For example, to launch a viral marketing campaign for a new product via Twitter, a known strategy is for the marketer to seed the product with a few selected influential users who can potentially influence a disproportionately large number of others and also quickly trigger a cascade of influence. As another example, in the event of a national crisis, the governmental authority can conduct a massive information campaign by disseminating truthful information via influential users to effectively achieve strategic goals and also counteract rumors. As the last example, to have realtime situational awareness about a physical region of interest, military agencies can recruit volunteers in the target region via influential Twitter users there and then outsource the collection of in-situ information to the volunteers.

The strong promise of influential users leads to the growing attention on how to measure the influence of a Twitter user \cite{ChaMea10,KwakWha10,WengTwi10,BakshEve11}. There are also over 20 commercial tools available for measuring twitterers' online influence. Common to these research proposals \cite{ChaMea10,KwakWha10,WengTwi10,BakshEve11} and commercial tools is to capture the qualitative feature of online influence as ``the ability to cause effect, change behavior, and drive measurable outcomes online'' \cite{SolisRis12} and to quantify a twitterer's online influence based on his/her \emph{interactions} with others.

The rise of \emph{social bot}s \cite{FerraRis14} or \emph{sybil}s \cite{DouceSyb02} in general on Twitter is jeopardizing trustworthy influence measurement. In a sybil attack, the adversary coordinates many fake accounts (also called \emph{bot}s or \emph{sybil user}s hereafter) to unfairly overpower non-sybil users. Despite various efforts to detect sybil users on Twitter \cite{BenevDet10,GrierSpa10,StrinDet10,ThomaDes11,GaoTow12,ThomaTra13}, sybil users are still thriving on Twitter. For example, a recent study \cite{SybilUnd12} revealed that at least 10\% of Twitter users are sybil users. Given the exclusive reliance of existing influence measurement techniques on user interactions, the adversary could coordinate his sybil users to create arbitrary interactions to inflate their influence scores on Twitter. Since influence scores are relatively defined, the adversary could also effectively deflate the influence scores of non-sybil Twitter users. According to our recent study \cite{ZhangOn13}, an adversary controlling 1,000 sybil users can quickly generate an influence score in the 95th percentile for any sybil user under popular influence measurement tools such as Klout \cite{Klout}, Kred \cite{Kred}, and Retweet Rank \cite{retweetrank}. In a similar study \cite{MessiYou13}, Messias \emph{et al.} used two social bots to successfully obtain high Klout scores.

The lack of sybil-resilient influence measurement services on Twitter can be detrimental. Specifically, there is a growing market for influence measurement services with more than 20 service providers available \cite{SolisRis12}. If these service providers fail to provide trustworthy measurement results due to sybil attacks, they will have extreme difficulty getting customers and surviving, and their customers could not achieve effective information dissemination or acquisition as expected.

The root cause for the vulnerability of existing influence measurement techniques to sybil attacks lies in the incautious use of user interactions. Specifically, Twitter permits four types of publicly visible user interactions, including \emph{follow}, \emph{retweet}, \emph{reply}, and \emph{mention}. The interactions about any user can be further classified into \emph{incoming} interactions towards him and \emph{outgoing} interactions from him. Since a sybil user can freely follow, retweet, reply to, and mention other sybil or non-sybil users, extensive outgoing interactions are fairly easy to create and thus unsuitable for sybil-resilient influence measurement. In addition, since sybil users could easily get many legitimate followers \cite{GhoshUnd12,YangAna12,StrinFol13}, the number of followers each user has should also be ruled out. In contrast, we observe from real Twitter data that non-sybil users tend to be more selective in retweeting, replying to, and mentioning other users. This observation is in line with the real-life scenario: one may exchange business cards with many strangers but will be more cautious in choosing whom to further interact with. This means that incoming retweets, replies, and mentions are much more trustworthy information for measuring user influence. Existing influence measurement techniques, however, use all incoming and outgoing interactions in a non-discriminative way.

We propose TrueTop, a novel sybil-resilient influence measurement system based on the incoming retweets, replies, and mentions each Twitter user has. TrueTop provides on-demand influence measurement services to various customers such as business companies and government agencies. Given a target set of Twitter users (e.g., those in a geographic area of interest), TrueTop outputs a ranked list of top-$K$ influential users for a desirable integer $K\geq 1$. TrueTop is designed to be \emph{sybil-resilient} and also \emph{accurate}, which means that the TrueTop output contains bounded sybil users and the true top-$K$ non-sybil users with overwhelming probability, respectively.

The main design challenge for TrueTop is that sybil users can arbitrarily interact among themselves, so it is not sybil-resilient to evaluate a user's influence directly based on his total incoming retweets, replies, and mentions. We propose the following method to tackle this challenge. Given the target set of users, we first construct a weighted directed \emph{interaction graph}, in which every vertex corresponds to a unique user in the target set. An edge from vertex $a$ to vertex $b$ exists if user $a$ has ever retweeted, replied to, or mentioned user $b$, and the edge weight is proportional to the number of retweets, replies, and mentions from $a$ to $b$. Imagine that the interaction graph consists of a virtual non-sybil region with all non-sybil users and a virtual sybil region with all sybil users. Given our previous observations, both the number of edges and the total edge weights from the non-sybil region to the sybil region should be much smaller than those in the reverse direction. Then we seed some carefully chosen vertices (or users) in the non-sybil region with some \emph{credits} and let every vertex in the whole graph allocate its current credits to its direct successors proportionally to the corresponding edge weights in every iteration. After sufficient iterations, the top-$K$ influential non-sybil users are very likely to stand out, as they can accumulate many credits due to their abundant incoming retweets, replies, and mentions. In contrast, the total credits flowing into the sybil region can be very limited, so even the sybil users with many incoming interactions from sybil followers may end up with few credits. We can thus achieve sybil-resilient influence measurement by counting the final credits at every vertex.

This paper makes the following contributions.

\begin{itemize}
\item We motivate and formulate the problem of sybil-resilient influence measurements on Twitter.

\item We propose TrueTop, a novel influence measurement system that can identify the top-$K$ influential users in a target set of Twitter users with high accuracy in the presence of sybil attacks by exploiting the selectivity of non-sybil users in interacting with other users.

\item We confirm the high accuracy and sybil-resilience of TrueTop by detailed theoretical analysis and extensive experiments on real datasets.

\end{itemize}

The rest of this paper is organized as follows. Section~\ref{sec:back} surveys the related work. Section~\ref{sec:problem} introduces Twitter basics, our system and threat models, and our design objectives. Section~\ref{sec:design} illustrates the TrueTop design. Section~\ref{sec:Analysis} theoretically analyzes the accuracy and sybil resilience of TrueTop. Section~\ref{sec:evaluation} evaluates the performance of TrueTop by detailed experiments. Section~\ref{sec:conclusion} concludes the paper.
\section{Related Work} \label{sec:back}
There is significant effort to explore social networks for effective sybil defenses in various distributed systems, such as SybilGuard \cite{YuSyb06} and SybilLimit \cite{YuSyb10} for P2P networks, SumUp \cite{TranSyb09} for online voting systems, and SybilInfer \cite{DanezSyb09}, SybilDefense \cite{WeiSyb12}, and SybilRank \cite{CaoAid12} for online social networks. A common assumption is that each node can be mapped into one in an undirected social network graph where every edge corresponds to a human-established trust relation. Although the attacker can create many sybil accounts, he cannot establish an arbitrarily large number of social trust relations with non-sybil users. Moreover, all schemes assume that the honest region is fast mixing and separate from the sybil region. Built upon these two key insights, these schemes conduct varying community detection methods \cite{ViswaAna11} to limit the number of sybil users admitted into or their impact in various application scenarios.

Recent measurement studies have questioned these two assumptions. Yang \emph{et al.} \cite{YangUnc14} showed that sybil users on the Facebook-like Renren network can have their friend requests accepted by many non-sybil users. A similar result targeting Facebook was reported in \cite{BoshmSoc11}. Blending sybil users into the non-sybil community would reduce the effectiveness of the existing sybil defenses \cite{KollSta14}. In addition, the work in \cite{RatkiTru11,GhoshUnd12, MessiYou13, ZhangOn13, FerraRis14} showed that sybil users successfully acquired a number of followings from non-sybil users on Twitter. All these findings indicate that neither bidirectional friendships in Fackbook-like OSNs nor unidirectional followings in Twitter-like microblogging systems can be used as the trustable mirroring of real social relations. Moreover, it has been shown in \cite{MohaiMea10, MohaiOn12} that the mixing time of many practical and directed social graphs is much longer than previously expected. Since neither of the two key assumptions underlying the schemes in \cite{YuSyb06,TranSyb09,DanezSyb09,ViswaAna11,YuSyb10,WeiSyb12, CaoAid12} holds in directed networks such as Twitter, they are not directly applicable to our targeted scenario. Our TrueTop system does not rely on either assumption.

As a special kind of sybil users, spammers in Twitter has attracted considerable attention in recent years. A common approach adopted by existing work \cite{BenevDet10,StrinDet10,GrierSpa10,ThomaDes11,GaoTow12,ThomaTra13,EgeleCOM13,HuSoc13} is to detect spammers by measuring the behavioral difference between spammers and legitimate users. Spammers are a special type of sybil users, and the detection of general sybil users on Twitter remains an open challenge.

There is a rich literature for influence measurement on Twitter. Cha \emph{et al.} \cite{ChaMea10} found that the numbers of retweets and mentions serve as better metrics than the number of followers in measuring user influence. Bakshy \textit{et al.} \cite{BakshEve11} proposed to measure user influence based on his ability to post the tweets that generates a cascade of retweets. TwitterRank \cite{WengTwi10} combines link structure and topical similarity between Twitter users and uses a modified PageRank algorithm to calculate user influence. Pal and Counts \cite{PalIde11} also proposed a framework to identify topical authorities in microblogging systems. All these schemes are vulnerable to sybil users who can forge arbitrary information employed by these schemes for influence measurement. Moreover, many metrics used by these schemes have been incorporated into commercial influence measurement tools \cite{SolisRis12}, and the vulnerability of representative tools to sybil attacks has been experimentally verified in \cite{ZhangOn13}.

Also related is the research on modelling, measuring, and analyzing the interactions in OSNs, e.g., \cite{ChunCom08,BenevCha09,WilsoUse09,JiangUnd13,XieInn12}. Our work is the first to build a weighted directed interaction graph from historical incoming retweets, replies, and mentions on Twitter and use it for identifying influential users.

\section{Preliminaries} \label{sec:problem}
\subsection{Twitter Basics}

We illustrate the basic operations on Twitter to help understand our design. The social relationships on Twitter are unidirectional by users \emph{following} others. If user $A$ follows user $B$, $A$ is $B$'s \textit{follower}, and $B$ is $A$'s \textit{followee}. A user usually needs no prior consent from his followees. Twitter also allows each user to approve/deny every following request, but this option is relatively rarely used. A user can send text-based messages of up to 140 characters, known as \emph{tweet}s, which can be read by all his followers. Tweets can be visible to anyone with or without a Twitter account, and they can also be protected and are only visible to approved followers. There are three special kinds of tweets corresponding to three operations. A \emph{retweet} is a re-posting of someone else's tweet, a \emph{reply}  corresponds to a response to a tweet, and a \emph{mention} refers to inserting ``@username'' in a tweet to ensure that the specified user can see this tweet. Finally, each user has a \textit{timeline} which shows all the latest tweets (including original tweets, retweets, replies, and mentions) of his followees. Also note that Twitter allows direct messages to be sent between users. Since those direct messages are not publicly visible, they cannot be used to measure user influence.

\subsection{System Model}

TrueTop is run by a service provider (SP) offering on-demand influence measurement services to customers such as viral marketers, government/military agencies, or even individuals. Given a measurement request, the TrueTop SP first determines the target set of Twitter users to evaluate, denoted by $\mathcal{U}$. The users in $\mathcal{U}$ can be directly given by the customer or identified by the TrueTop SP according to some common features specified by the customer. For example, the customer can specify a target geographic region, a target age group, a target topic (e.g., music), etc. As said, TrueTop relies on incoming interactions among the users in $\mathcal{U}$, i.e., the retweets, replies, and mentions each user in $\mathcal{U}$ has received from all the other users in $\mathcal{U}$. So we assume that the SP has a reliable way to obtain the incoming interaction data needed, e.g., directly from Twitter, via crawling, or from some third-party providers of social media data. For example, Gnip (\url{http://gnip.com/}) is an authorized reseller of Twitter data. TrueTop is designed to output a ranked list of top-$K$ influential users in $\mathcal{U}$, where $K\geq 1$ denotes a customer-specified integer.

\subsection{Threat Model}
Let $\tilde{\mathcal{U}}$ denote all possible sybil users in $\mathcal{U}$. We assume that the SP knows neither which user in $\mathcal{U}$ is a sybil user nor how many sybil users there are; otherwise, the identified sybil users can be simply removed from $\mathcal{U}$. Based on the recent measurement study \cite{GhoshUnd12}, we assume that each sybil user may have followed and also been followed by some non-sybil and sybil users in $\mathcal{U}$. There may be a single attacker controlling $\tilde{\mathcal{U}}$ or multiple independent ones with each controlling an exclusive subset of $\tilde{\mathcal{U}}$. TrueTop can deal with both cases without modification, so we focus on the more challenging former case hereafter. The goal of the attacker is to gain high influence scores for his sybil users and maximize the number of users in the TrueTop output.

\subsection{Design Objectives}
Let $\mathcal{U}_K^\ast$ and $\mathcal{U}_K$ denote the top-$K$ non-sybil influential users in $\mathcal{U}$ and the TrueTop output, respectively. We have two major design objectives.
\begin{itemize}
\item \emph{Accuracy}: TrueTop should identify the true top-$K$ non-sybil users, which means the difference between $\mathcal{U}_K^\ast$ and $\mathcal{U}_K$ should be very small.
\item \emph{Sybil resilience}: TrueTop should not identify sybil users as top-$K$ users, i.e., the the intersection $\mathcal{U}_K\cap \tilde{\mathcal{U}}$ should be very small.
\end{itemize}

\section{TrueTop design}\label{sec:design}

\subsection{Overview}

TrueTop is motivated by the observation that incoming retweets, replies, and mentions are more trustworthy for measuring user influence than outgoing interactions. So our first step is to construct an interaction graph, in which every vertex corresponds to a unique user in the target set $\mathcal{U}$, and every directed edge indicates totally non-zero retweets, replies, and mentions from the tail user to the head user. In addition, the weight of every edge is a non-decreasing function of related retweets, replies, and mentions.

The next step is to choose a suitable metric to quantify the influence of every user (vertex) in the interaction graph. TrueTop adopts \emph{weighted eigenvector centrality} (WEC for short) \cite{NewmaAna04}, the de facto metric for measuring the influence of a node in a weighted directed graph. Specifically, the WEC score of every user corresponds to his influence score, which depends on the weights of his incoming edges, the number of his direct predecessors, and their influence scores which are further determined by their respective incoming edges and direct predecessors. The WEC score reflects an intuition that the influence of a user is better indicated by the interactions from influential users than those from less influential users.

We uses iterative credit distribution for the convenience to describe and understand our method. Specifically, we select some random users (called \emph{seed}s) in the interaction graph and seed each with some \emph{credit}s. In each iteration, we allocate all the credits each user receives in the last iteration to his direct successors proportionally to individual edge weights. The credits each user receives in one iteration are expected to stabilize after sufficient iterations and be proportional to his WEC score. It can be easily shown that iterative credit distribution is equivalent to power iteration \cite{LangvDee04}, a standard technique for computing WEC scores. Since sybil users can create arbitrary interactions among themselves, some of them may gain enough credits to appear in the top-$K$ list. TrueTop achieves high sybil resilience by carefully choosing the initial seeds and also early terminating iterative credit distribution.

In what follows, we first illustrate the construction of the interaction graph in Section~\ref{sec:Graph}. Next, we present an iterative credit distribution scheme over the interaction graph in Section~\ref{sec:Credit}. Finally, we introduce how to achieve sybil-resilient iterative credit distribution in Section~\ref{sec:Early}.

\subsection{Interaction Graph Construction}\label{sec:Graph}

Given the target users $\mathcal{U}$ and their interaction data, TrueTop first builds a weighted directed interaction graph denoted by $\mathcal{G}=\langle \mathcal{U}, \mathcal{V}\rangle$, where $\mathcal{U}$ is abused to denote the vertex set, and every edge $v_{i,j}\in \mathcal{V}$ ($i,j\in \mathcal{U}$) is directed and indicates that there are some retweets, replies, and/or mentions from user $i$ to $j$. The major challenge here is to determine the weight $w_{i,j}$ of every edge $v_{i,j}$. As shown in Fig.~\ref{fig:interaction}, $\mathcal{G}$ can be divided into a virtual sybil region $\mathcal{S}$ including all the sybil users and a virtual non-sybil region $\mathcal{H}$ including all the non-sybil users. The sybil-resilience requirement for TrueTop requires that the sum of the edge weights from the non-sybil region to the sybil region is small, while the accuracy requirement for TrueTop demands that the weight $w_{i,j}$ reflects the true influence of user $j$ on $i$ in the target period. Let $\mathcal{I}_{i,j}$ denote the set of time-indexed retweets, replies, and mentions from user $i$ to $j$. We consider the following two methods for defining the edge weights.
\begin{itemize}
\item \emph{Sum-based.} In this method, $w_{i,j}$ equals $|\mathcal{I}_{i,j}|$. Sum-based edge weights satisfy the sybil-resilient requirement, as the total edge weights from the non-sybil region to the sybil region are as limited as the number of retweets, replies, and mentions from non-sybil users to sybil users. They also partially satisfy the accuracy requirement, as the more interactions from $i$ to $j$, the more influence $j$ likely has on $i$, and the higher $w_{i,j}$. Sum-based edge weights, however, fail to catch the temporal aspect of interactions. For example, consider another direct predecessor of $j$, say $l$, where $|\mathcal{I}_{i,j}|=|\mathcal{I}_{l,j}|$. Assume that the interactions in $\mathcal{I}_{l,j}$ occurred in the last few days in the target period, while those in $\mathcal{I}_{i,j}$ were spread more evenly. It may be natural to say that $j$ has stronger influence on user $i$ than on user $l$, but we have $w_{i,j}= w_{l,j}$ for sum-based methods.

\item \emph{Entropy-based.} In this method, we divide the target period into $\mu$ equal-length epochs for some system parameter $\mu\geq 1$ and denote the total number of retweets, replies, and mentions from user $i$ to $j$ in epoch $x$th by $d_x$, where $|\mathcal{I}_{i,j}|=\sum_{x=1}^\mu d_x$. Then we define the edge weight $w_{i,j} = (1-\sum_{x=1}^\mu \frac{d_x}{|\mathcal{I}_{i,j}|} \log \frac{d_x}{|\mathcal{I}_{i,j}|}) |\mathcal{I}_{i,j}|$. The more consistent the interactions from $i$ to $j$ in time, the higher $w_{i,j}$, and vice versa. When all the interactions happen in a single epoch, the weight is identical to sum-based $|\mathcal{I}_{i,j}|$. Entropy-based edge weights can also satisfy the sybil-resilience requirement, as non-sybil users unlikely have consistent interactions to sybil users so that the total edge weight from the non-sybil region to the sybil region can be expectedly small. In contrast to sum-based edge weights, entropy-based edge weights successfully catch the temporal information in the interactions while failing to reflect the volume of the interactions. So they partially satisfy the accuracy requirement as well.
\end{itemize}
The effects of the above methods are compared in Section~\ref{sec:evaluation}. There may be other ways to define the edge weights. For example, we can let $w_{i,j}$ equal a linear combination of the edge weights derived under sum-based and entropy methods, respectively; we can also assign different weights to retweets, replies, and mentions according to slightly different effort and/or social implication related to performing these interactions. A further study on such issues is left as future work due to space constraints.

Note that we only consider retweets, replies, and mentions in the weight definitions because they are representative on Twitter and have been used in all the existing influence measurement techniques. Some other factors could also impact the user influence, such as following connections and favorites. As stated before, since sybil users could easily get many legitimate followers \cite{GhoshUnd12,YangAna12,StrinFol13}, the following connections fail to achieve the sybil resilience and hence should be ruled out for the influence measurement. On Twitter, a user could favor the tweets from other users, but there is no public Twitter API which can return the favorite user list for any given tweet. Should a public Twitter API for retrieving favorites become available, we can easily incorporate favorites into TrueTop.

\subsection{Credit Distribution}\label{sec:Credit}

TrueTop uses the WEC score of every user in $\mathcal{G}=\langle \mathcal{U}, \mathcal{V}\rangle$ as his influence score. Specifically, let $\pi_i$ denote the WEC score of user $i$ in $\mathcal{G}$ and ${\bf W}=(w_{i,j})$ denote the normalized weighted adjacency matrix of $\mathcal{G}$. The vector ${\boldsymbol \pi} = \langle \pi_1,\pi_2,\dots,\pi_{|\mathcal{U}|}\rangle$ is the dominant eigenvector of ${\bf W}$, i.e., the solution to the equation ${\boldsymbol \pi} {\bf W} = {\boldsymbol \pi}$ according to \cite{NewmaAna04}.

Power iteration \cite{LangvDee04} is a common technique to compute the WEC vector ${\boldsymbol \pi}$. Let ${\bf v}_0$ be a random vector composed of $|\mathcal{U}|$ nonnegative elements totalling one. In power iteration, ${\boldsymbol \pi}$ is computed in an iterative fashion as
\begin{equation}\label{eq:powerEigen}
{\boldsymbol \pi} = \lim_{t\rightarrow \infty}{\bf x}^{(t)} = \lim_{t\rightarrow \infty} {\bf v}_0 {\bf W}^t\;,
\end{equation}
where ${\bf x}^{(t)} = {\bf x}^{(t - 1)} {\bf W}$ with the initial ${\bf x}^{(0)} = {\bf v}_0$. If $\mathcal{G}$ is strongly connected, $\boldsymbol \pi$ exists, is unique, and is unrelated to ${\bf v}_0$. In practice, power iteration normally terminates if $\|{\bf x}^{(t)} -{\bf x}^{(t-1)}\|_1 < \nu$ for some acceptable error threshold $\nu$ (e.g., $10^{-9}$).

The WEC vector only exists in a strongly connected graph \cite{NewmaAna04}, in which every vertex is reachable from every other vertex. Although $\mathcal{G}$ itself may be not strongly connected in practice, it usually has a giant strongly connected component (GSCC) which includes the majority of the vertexes and edges and is dramatically larger than all other strongly connected components (SCCs). Since the most influential users should have intensive interactions with other users, the top-$K$ influential users should be in the GSCC with overwhelming probability. Our subsequent operations thus apply to the GSCC only. The verification of the existence of GSCC in real datasets is deferred to Section \ref{sec:evaluation}.

TrueTop uses iterative credit distribution instead to compute ${\boldsymbol \pi}$ to facilitate the presentation. Initially, we randomly select a few users (called \emph{seed}s) in $\mathcal{G}$ and initialize each with the same number of notional credits totalling one. At every iteration, we allocate the credits each user receives in the last iteration to his direct successors proportionally to the corresponding edge weights. Let $C^{(t)}_j$ denote the number of credits at any user $j\in \mathcal{U}$ after $t$ iterations, which are proportional to his influence score measured after $t$ iterations. $C^{(t)}_j$ is a real number in general and can be computed as
\begin{equation}\label{eq:credit}
C^{(t)}_j = \sum_{i \in \mathtt{IN}(j)} \frac{w_{i,j}C^{(t-1)}_i}{\sum_{k\in \mathtt{OUT}(i)}w_{i,k}},
\end{equation}
where $\mathtt{IN}(j)$ and $\mathtt{OUT}(i)$ denote the direct predecessors of user $j$ and the direct successors of user $i$ in $\mathcal{G}$, respectively. Similarly, we can terminate credit distribution when $\sum_{j\in \mathcal{U}} |C^{(t)}_j - C^{(t-1)}_j|< \eta$ for some acceptable error threshold $\eta$ (e.g., $10^{-9}$).

We can easily show that iterative credit distribution above is equivalent to power iteration. In particular, assume that $s$ seeds are chosen in iterative credit distribution, each having $1/s$ credits initially. We further select ${\bf v}_0$ for power iteration such that the $i$th element equals $1/s$ if user $i$ is a seed and zero otherwise. Then Eq.~(\ref{eq:credit}) is apparently the element-wise expression of ${\bf x}^{(t)} = {\bf x}^{(t - 1)} {\bf W}$. Since power iteration does not depend on a specific ${\bf v}_0$, we have $x^{(t)}_j = C^{(t)}_j$ for any user $j\in \mathcal{U}$ after $t$ iterations.

\begin{figure}
\centering
\includegraphics[width=0.5 \textwidth]{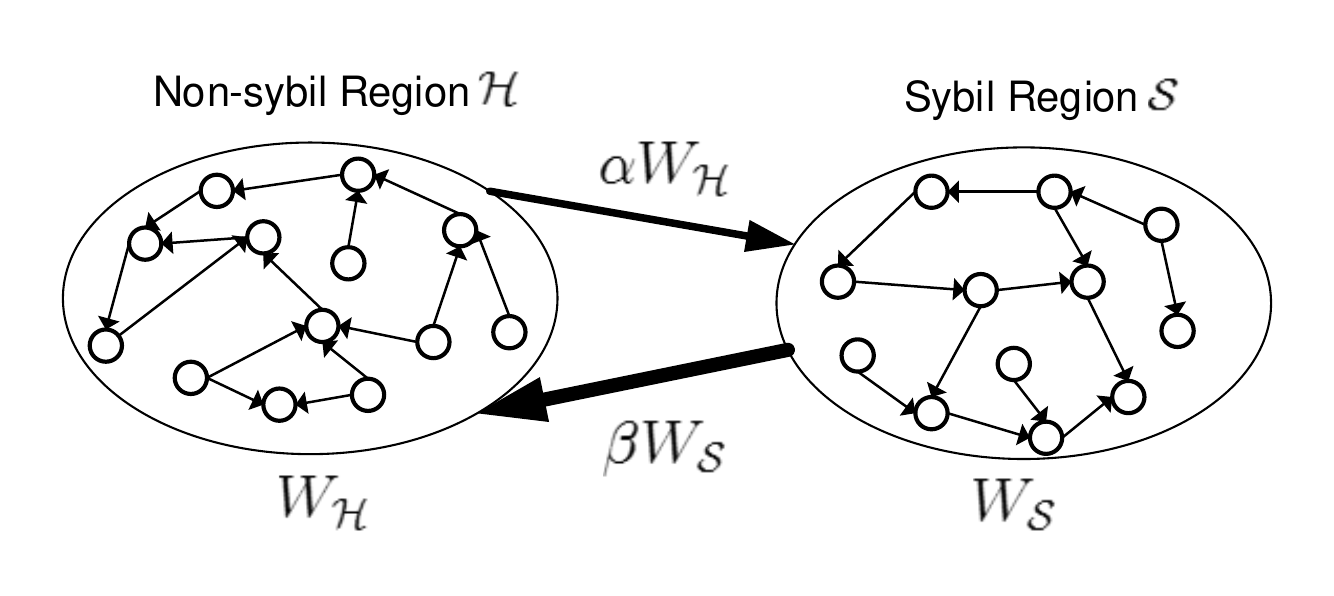}
\caption{The interaction graph with a virtual non-sybil region $\mathcal{H}$ and a virtual sybil region $\mathcal{S}$. }
\label{fig:interaction}
\vspace{-0.3in}
\end{figure}

Iterative credit distribution described above is still subject to sybil attacks. To see this, consider Fig.~\ref{fig:interaction} where the interaction graph is divided into a virtual non-sybil region $\mathcal{H}$ and a virtual sybil region $\mathcal{S}$. We denote the total edge weights within $\mathcal{H}$, within $\mathcal{S}$, from $\mathcal{H}$ to $\mathcal{S}$, and from $\mathcal{S}$ to $\mathcal{H}$ by $W_\mathcal{H}$, $W_\mathcal{S}$, $\alpha W_\mathcal{H}$, and $\beta W_\mathcal{S}$, respectively, where $\alpha\ll 1$. Although the adversary has no control over $W_\mathcal{H}$ and $\alpha$, he can easily manipulate $W_\mathcal{S}$ and $\beta$ to make $\beta W_\mathcal{S}$ very small. Even if all the seeds are chosen from $\mathcal{H}$ in the best scenario, more and more credits will flow into and stay in $\mathcal{S}$ as time goes by. We have the following proposition about the vulnerability of iterative credit distribution to sybil attacks.
\begin{proposition}
Assume that the total edge weights from the non-sybil region $\mathcal{H}$ to the sybil region $\mathcal{S}$ and from $\mathcal{S}$ to $\mathcal{H}$ are $\alpha$ and $\beta$ fractions of the total edge weights in $\mathcal{H}$ and $\mathcal{S}$, respectively. The total credits in $\mathcal{S}$ increase monotonically with the iteration $t$ and asymptotically approach to $\frac{\alpha}{\alpha+\beta}$. \label{th:credit}
\end{proposition}
The proof of Proposition~\ref{th:credit} can be found in Appendix~I-A. Since the adversary can well control the topology within $\mathcal{S}$, most credits in $\mathcal{S}$ can go to a few sybil users who may eventually appear in the top-$K$ influential users.

\subsection{Sybil-Resilient Credit Distribution}\label{sec:Early}

TrueTop adopts the following two defenses against sybil attacks such that most credits can stay in the non-sybil region for sufficient iterations.

The first defense is to use non-sybil seeds only so that credit distribution can start from the non-sybil region $\mathcal{H}$. We propose to use verified Twitter users as seeds by three reasons. First, Twitter has certified their authenticity. Each verified user has a blue verified badge on his profile page and is followed by the official Twitter account \emph{@verified}. Second, there are many verified users available as candidate seeds. As of April 2014, Twitter has verified more than 88,600 accounts among 255 million monthly active users and keeps verifying more. Since $\mathcal{G}$ can be expected to contain many users in practice, there should be at least one verified user in $\mathcal{G}$ with very high probability. Finally, since verified users are usually public figures such as politicians, celebrities, or business leaders,  we can trust them to be very cautious in whom to retweet, reply to, and mention. This implies that the immediate successors of verified users on the interaction graph $\mathcal{G}$ are very likely to be non-sybil users as well, so are the successors' immediate successors. If we start credit distribution from verified users, most credits can be expected to stay inside $\mathcal{H}$ after many iterations.

How many seeds should we choose? Some verified users may be very close to the sybil region, but we cannot tell who they are. Ideally speaking, we should choose the verified users far from the sybil region. On the one hand, if a verified user is randomly chosen as the sole seed, he may be too close to the sybil region. On the other hand, if we use all the verified users in $\mathcal{G}$ as the seeds, it is very likely that some of them are close to the sybil region. In addition, the number of seeds affects the convergence of iterative credit distribution: the more seeds, the faster the algorithm converges. It is impossible to specify the decisive rules for seed selection, so we randomly choose $s\geq 1$ seeds from the verified users in $\mathcal{G}$ and experimentally evaluate the impact of seed selection in Section~\ref{sec:evaluation}.

How should we assign the initial credits among the $s$ seeds? We propose two methods as follows.
\begin{itemize}
\item \emph{Basic method}. The total credits are evenly assigned to the $s$ seeds. This straightforward method assumes that each seed has the same importance for credit distribution.
\item \emph{Reverse-WEC}. Since the credits flow out from the seeds, we can assign more initial credits to the seeds who can quickly reach more users to speed up the algorithm convergence. For this purpose, we conduct the credit distribution introduced in Section~\ref{sec:Credit} over an inverse interaction graph generated from $\mathcal{G}$ by reversing the directions of all the edges and also setting all the edge weights to one. The final credits at each user naturally reflects his connectivity in $\mathcal{G}$. So
    we select the verified users with the top-$K$ highest credits as the seeds and then assign to each of them the initial credits proportional to their credits obtained via reverse credit distribution.
\end{itemize}

The second defense is to early terminate iterative credit distribution before it converges in the whole graph $\mathcal{G}$. To see the necessity and intuition for this defense, recall that we start credit distribution from non-sybil seeds in the non-sybil region. Since the total edge weight from the non-sybil region to the sybil region is relatively small, we can expect credit distribution to converge much faster in the non-sybil region than in the whole $\mathcal{G}$. In addition, the most influential non-sybil users normally have many incoming interactions and thus a rich number of credit sources in $\mathcal{G}$. So they can quickly accumulate a lot of credits to stand out much faster than other non-sybil users. If we early terminate iterative credit distribution, most or all of the sybil users would not get enough credits to appear in the resulting top-$K$ influential users, so we can achieve sybil resilience. However, if credit distribution stops too early, some true top-$K$ influential non-sybil users may not get enough credits to be ranked in the top-$K$ list, leading to an inaccurate result.

We design a simple but effective algorithm to tackle the dilemma between sybil resilience and accuracy. The key idea is to monitor the ranking change of the candidate top-$K$ users in two consecutive iterations. Whenever the ranking change is no larger than an acceptable threshold, we terminate the algorithm and output the current top-$K$ users as the top-$K$ influential users. This algorithm is directly built on our observation above. Specifically, since the top-$K$ non-sybil influential users is more likely to stand out much faster than both sybil users and other non-sybil users during credit distribution, their rankings are more likely to become stable in fewer iterations as well. We detail the algorithm as follows and postpone its performance analysis to Section~\ref{sec:Analysis}.

\begin{algorithm}
 \SetKwInOut{Input}{input}\SetKwInOut{Output}{output}
\caption{Find the top-$K$ influential users}
 \Input{Interaction graph $\mathcal{G}$; $s$ seed users; $K$; maximum number of iterations $T$; ranking-error tolerance $\epsilon$}
 \Output{The top-$K$ influential users}

 Assign initial credits among $s$ seed users by either basic or reverse-WEC method\;
 $t\leftarrow 1$\;
 \While{$t < T$}{
    Distribute the credit in the $t$-th iteration according to Eq.~\ref{eq:credit}\;
    Rank the users by their credits and obtain the candidate top-$K$ users $\mathcal{R}^{(t)}$\;
    Compute the ranking distance $d(K)^{(t)}$ between $\mathcal{R}^{(t)}$ and $\mathcal{R}^{(t-1)}$ as in Eq.~\ref{eq:distance}\;
    \If{$d(K)^{(t)} <= \epsilon$ }
    {\textbf{break}\;}
    $t\leftarrow t+1$\;
 }
 \Return $\mathcal{R}^{(t)}$ as the top-$K$ influential users
 \label{alg:credit}
\end{algorithm}

 Let $r^{(t)}(u)$ and $r^{(t-1)}(u)$ denote the rankings of user $u$ in iterations $t$ and $t-1$, respectively. We define the ranking distance $d(K)^{(t)}$ between $\mathcal{R}^{(t)}$ and $\mathcal{R}^{(t-1)}$ as
\begin{equation}
d(K)^{(t)} = \sum_{u\in \mathcal{R}^{(t)}(K)\cup \mathcal{R}^{(t-1)}(K)} |r^{(t)}(u) - r^{(t-1)}(u)|\;. \label{eq:distance}
\end{equation}

The algorithm above has two key parameters: $T$ and $\epsilon$. The former dictates the maximum number of iterations, and the latter specifies the maximum ranking error tolerance. The larger $T$, the longer the algorithm execution time, the more accurate the top-$K$ influential users, the more credits flowing into the sybil region and thus the less sybil resilience, and vice versa. In contrast, the larger $\epsilon$, the shorter the algorithm execution time, the less accurate the top-$K$ influential users, the fewer credits flowing into the sybil region and thus the higher sybil resilience, and vice versa. In practice, we can let $\epsilon < K$, meaning that each user in the current top-$K$ list has experienced a ranking change of less than one on average in contrast to the previous iteration.

\section{Performance Analysis}\label{sec:Analysis}
In this section, we analyze the accuracy and sybil resilience of TrueTop. Recall that $\mathcal{U}_K^\ast$ denotes the true top-$K$ influential users in the non-sybil region, $\mathcal{U}_K$ denotes the TrueTop output (i.e., the output of Alg.~\ref{alg:credit}), and $\tilde{\mathcal{U}}$ denotes all the sybil users in the sybil region. So we can use $\mathcal{U}_K \cap\mathcal{U}_K^\ast$ and $\mathcal{U}_K\cap \tilde{\mathcal{U}}$ to measure the accuracy and sybil-resilience of TrueTop, respectively.

To make the performance analysis tractable, we first assume that Alg.~\ref{alg:credit} runs in the non-sybil region only, so we can conduct an upper-bound analysis about the accuracy of TrueTop by setting the ranking error tolerance parameter $\epsilon =0$ and $T$ extremely large such that Alg.~\ref{alg:credit} terminates only when a stable top-$K$ user list is found. We then show that Alg.~\ref{alg:credit} will terminate in asymptotically the same number of iterations for $\epsilon=0$, based on which we finally estimate the number of sybil users appearing in $\mathcal{U}_K$.  As stated before, the larger $\epsilon$, the shorter the algorithm execution time, the less accurate the top-$K$ influential users, the fewer credits flowing into the sybil region and thus the higher sybil resilience, and vice versa. Hence by setting $\epsilon=0$, we can provide the lower and upper bounds for sybil resilience and accuracy, respectively. As for arbitrary $\epsilon > 0$, we unfortunately cannot obtain the closed-form analytical result for sybil resilience or accuracy and thus resort to experiments to evaluate its impact in Section~\ref{sec:evaluation}.

The following concepts are needed for the accuracy analysis.
\begin{definition}[(Relative) Error Bound]
Let ${\boldsymbol \pi}$ denote the true WEC vector of non-sybil users and the $k$-ranked user refer to the one with the $k$th highest WEC score $\tau_k$ in ${\boldsymbol \pi}$. Let $\tau^{(t)}_k$ denote the WEC score of the $k$-ranked user after iteration $t$. Then $e^{(t)}_k=|\tau^{(t)}_k -\tau_k|$ is defined as the error bound for the $k$-ranked node after iteration $t$, and $e'^{(t)}_k = e^{(t)}_k / \tau_k$ is defined as the relative error bound.
\end{definition}
\begin{definition}[(Relative) WEC gap]
The WEC gap for the $k$-ranked node is defined as $\Delta_k = \tau_k -\tau_{k+1}$, and $\Delta'_k = \Delta_k / \tau_k$ is the correspondingly relative WEC gap.
\end{definition}

\begin{lemma}
Let $\bf W$ denote the normalized weighted adjacency matrix of the non-sybil region with $n$ users, among which there are $s$ seed users. Construct ${\bf v}_0$ for power iteration (see Eq.~\ref{eq:powerEigen}) such that the $i$th element equals $1/s$ if user $i$ is a seed and zero otherwise. Then the relative error bound for the $k$-ranked user satisfies $e'^{(t)}_k \leq \lambda^t $, where $\lambda < 1$ denotes ${\bf W}$'s second largest eigenvalue. \label{th:error}
\end{lemma}
Lemma~\ref{th:error} states that the rank of each user in iteration $t$ approaches its true rank for sufficiently large $t$. The proof of Lemma~\ref{th:error} can be found in Appendix~I-B.

In addition, Ghoshal and Barabasi \cite{GhoshRan11} recently found that if the WEC vector (Pagerank in their paper) follows power law distribution, the gap between the $k$th and $(k+1)$th WEC scores decreases with $k$. We thus have the following lemma.
\begin{lemma}\cite{GhoshRan11}
If the WEC vector ${\boldsymbol \pi}$ follows a power-law distribution with parameter $\gamma$, the relative WEC gap for the $k$-ranked user satisfies $\Delta'_k \approx \frac{1} {k(\gamma - 1)}$. \label{th:gap}
\end{lemma}
The proof of Lemma~\ref{th:gap} is straightforward according to \cite{GhoshRan11} and omitted here due to space constraints. In Section~\ref{sec:evaluation}, we show that the WEC vectors for real Twitter datasets indeed follow the power-law distribution. We then have the following theorem based on Lemma~\ref{th:error} and Lemma~\ref{th:gap}.
\begin{table*}[ht]
    \caption{Dataset Characteristics.}
    \centering
    \begin{tabular}{|c|c|c|c|c|}
       \hline
       & \texttt{SF} & \texttt{TS} & \texttt{Random} & \texttt{Music}\\
       \hline
       Crawling period & \multicolumn{2}{|c|}{8/30-11/30, 2013} & \multicolumn{2}{|c|}{6/28-9/28, 2013} \\
       \hline
       \# of users & 176,506 & 5,827 & 1,999,834 & 999,807 \\
       \# of edges & 1,493,924 & 40,031 & 63,803,204 & 34,688854\\
       \hline
       \# of users in GSCC & 104,000(58.9\%) & 4,127(70.8\%) & 1,541,343 (77.1\%) & 687,693 (68.9\%)\\
       \# of edges in GSCC & 1,305,834(87.4\%) & 36,189 (90.4\%) & 55,781,520 (87.4\%) & 30,170,774(87.0\%)\\
\# of users in the 2nd largest SCC & 357 & 6 & 82 & 21\\
        \hline
    \end{tabular}
    \label{tlb:size} \label{tlb:datasets}
\end{table*}

\begin{theorem}
For iterative credit distribution in a strongly-connected weighted directed graph with the monotone-decreasing $\Delta'_k$ with $k$, if $\lambda^t  \leq \Delta'_k/2$ in iteration $t$, the
ranked list of users with top-$k$ credits remain the same in subsequent iterations. \label{th:accuracy}
\end{theorem}
The proof is in Appendix~I-C. Theorem~\ref{th:accuracy} indicates that if there are no sybil users, Alg.~\ref{alg:credit} (or TrueTop) can generate the true top-$K$ influential non-sybil users if $\lambda^t \leq \Delta'_K / 2$, i.e., when $t \leq -\log(2K (\gamma - 1)) / \log(\lambda)$ or $ t = O(|\log(K) / \log(\lambda)|)$ iterations. This also corresponds to the case of $\epsilon=0$ with 100\% accuracy.  Since the total edge weights from/to the non-sybil region to/from the sybil region are relatively very small, we can expect that the sybil region has little impact on the influence rankings of non-sybil users. So the accuracy of TrueTop under sybil attacks is tightly related to how many sybil users can show up in the top-$K$ list, i.e., the sybil-resilience of TrueTop, as analyzed in the following theorem.

\begin{theorem}
Let $\alpha$ be the ratio of the total edge weight from the non-sybil region to the sybil region over the total edge weights in the non-sybil region. Assume that the attacker wants to place as many sybils into the top-$K$ list as possible by retaining all the credits flowing into the sybil region. The number of sybil users in the top-$K$ list after early termination in $t=O(\log(K) / \log(\lambda))$ iterations is upper-bounded by $K(1- (1- \alpha)^t) / (1- \alpha)^t$. \label{th:powerbound}
\end{theorem}
The proof is in Appendix~I-D. Accordingly, we can easily derive the lower bound for the accuracy of TrueTop because there are at least $K(2-1/(1-\alpha)^t)$ true top-$K$ non-sybil users in the final top-$K$ list. Note that since $\alpha \ll 1$ and $K$ is usually at the scale of 1,000 and 10,000, this upper bound is far less than $K$, meaning that there are only negligible sybil users in the top-$K$ list.

\section{Evaluation}\label{sec:evaluation}
In this section, we thoroughly evaluate the performance of TrueTop. We first introduce some implementation details and the runtime performance, followed by the datasets used in our evaluations. Next, we verify two underlying assumptions in our design. Finally, we evaluate the accuracy and sybil resilience of TrueTop under various sybil attacks.

\subsection{Implementation and Runtime Performance}\label{sec:Runtime}
TrueTop is composed of two main components: the interaction graph construction and the credit distribution with early termination. We implemented both with a total of 2000+ lines of mixed code of Python and C++. Specifically, to efficiently handle the large-scale interaction networks (millions of nodes and billions of edges) in a commodity PC, we adopted the Graphchi computing framework \cite{KyrolGra12} to implement the credit distribution of TrueTop. On our desktop with 3.4GHz Intel-i7 3770 CPU, 16G Memory, a 7200RPM hard disk, and Ubuntu 12.04 LTS, one single iteration of credit distribution took 0.3s, 2.5s, 9.2s, and 17.1s for our four datasets in Table \ref{tlb:datasets} with 4K, 10K, 1M and 2M nodes, respectively. For a graph with 2M nodes, TrueTop can thus find the top-1000 influential users after 1,000 iterations within less than five hours on a commodity PC. Since TrueTop is expected to be run by a service provider with much more powerful computation resources, its runtime performance should be acceptable.

\subsection{Datasets}\label{sec:Dataset}

We crawled four representative datasets with public Twitter APIs. The \texttt{SF} and \texttt{TS} datasets include all the active users who have specified San Francisco Bay Area and Tucson, Arizona in the location field of their public profiles in the crawling (or target) period, respectively. In addition, the \texttt{Random} dataset contains a random set of active Twitter users in the target period, and the \texttt{Music} dataset contains the active users who have used the keyword ``music'' in their tweets in the target period. Each dataset includes all the user IDs and also their time-indexed tweets during the target period, which include original tweets, retweets, replies, and mentions. Then we constructed two interaction graphs for each dataset according to the process in Section~\ref{sec:Graph}, one for sum-based edge weights and the other for entropy-based edge weights.

Table~\ref{tlb:datasets} summarizes the basic statistics of the interaction graphs of each dataset, which apply to both sum-based and entropy-based edge weights. As we can see, each interaction graph has a giant strongly connected component (GSCC) which is far larger than the second largest SCC. Since TrueTop measures user influence based on incoming interactions, the top-$K$ influential users are in the GSCC with overwhelming probability. Our subsequent evaluations are thus done on the GSCC in each interaction graph only. We obtained very similar evaluation results for sum-based and entropy-based interaction graphs. Due to space limitations, we report the results for sum-based interaction graphs in most cases.
\begin{figure}[t]
\centering
 \subfloat[Sum-based  \label{fig:dist-1}]{%
      \includegraphics[width=0.24\textwidth]{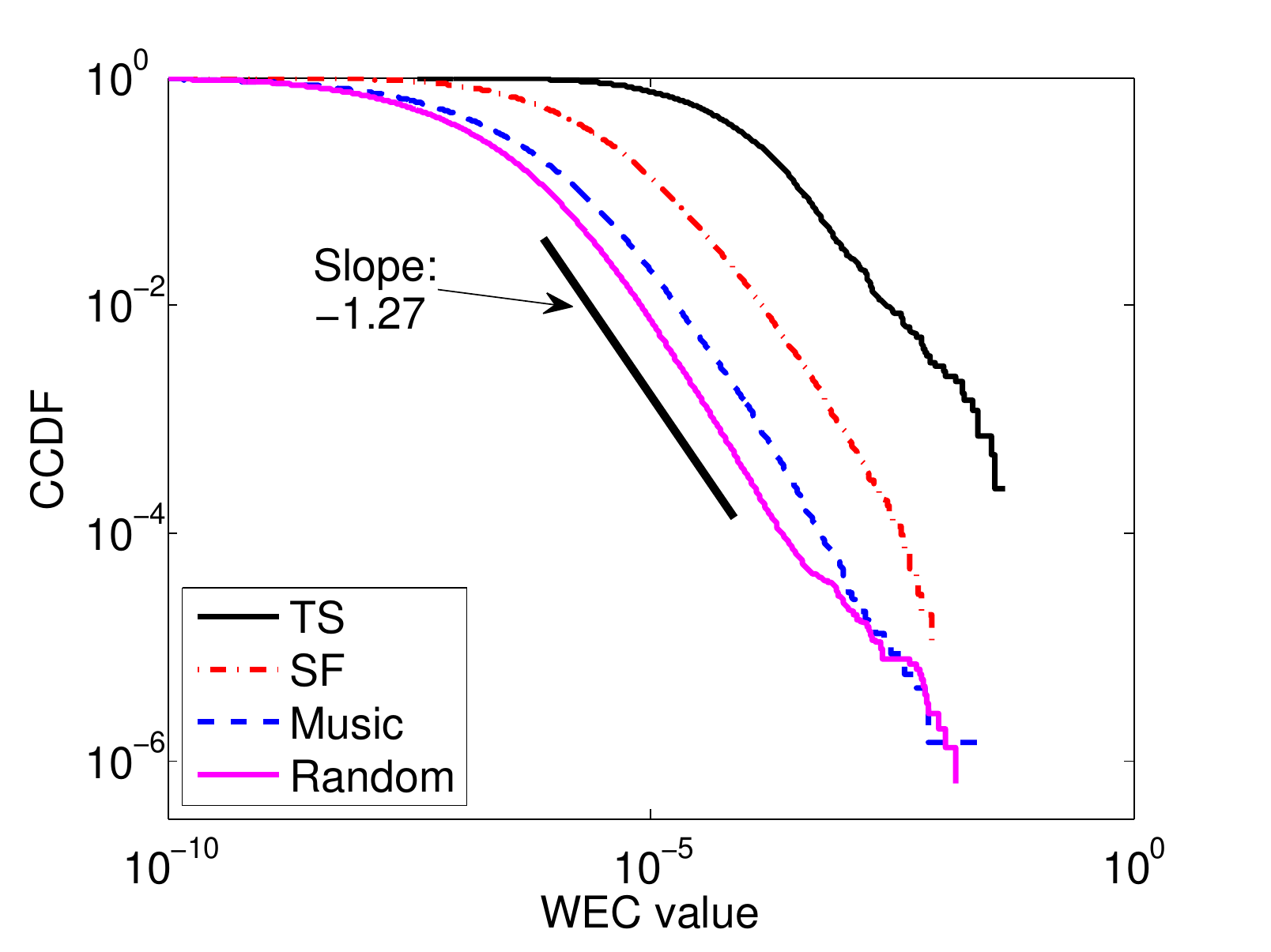}
    }
 \subfloat[Entropy-based \label{fig:dist-2}]{%
      \includegraphics[width=0.24\textwidth]{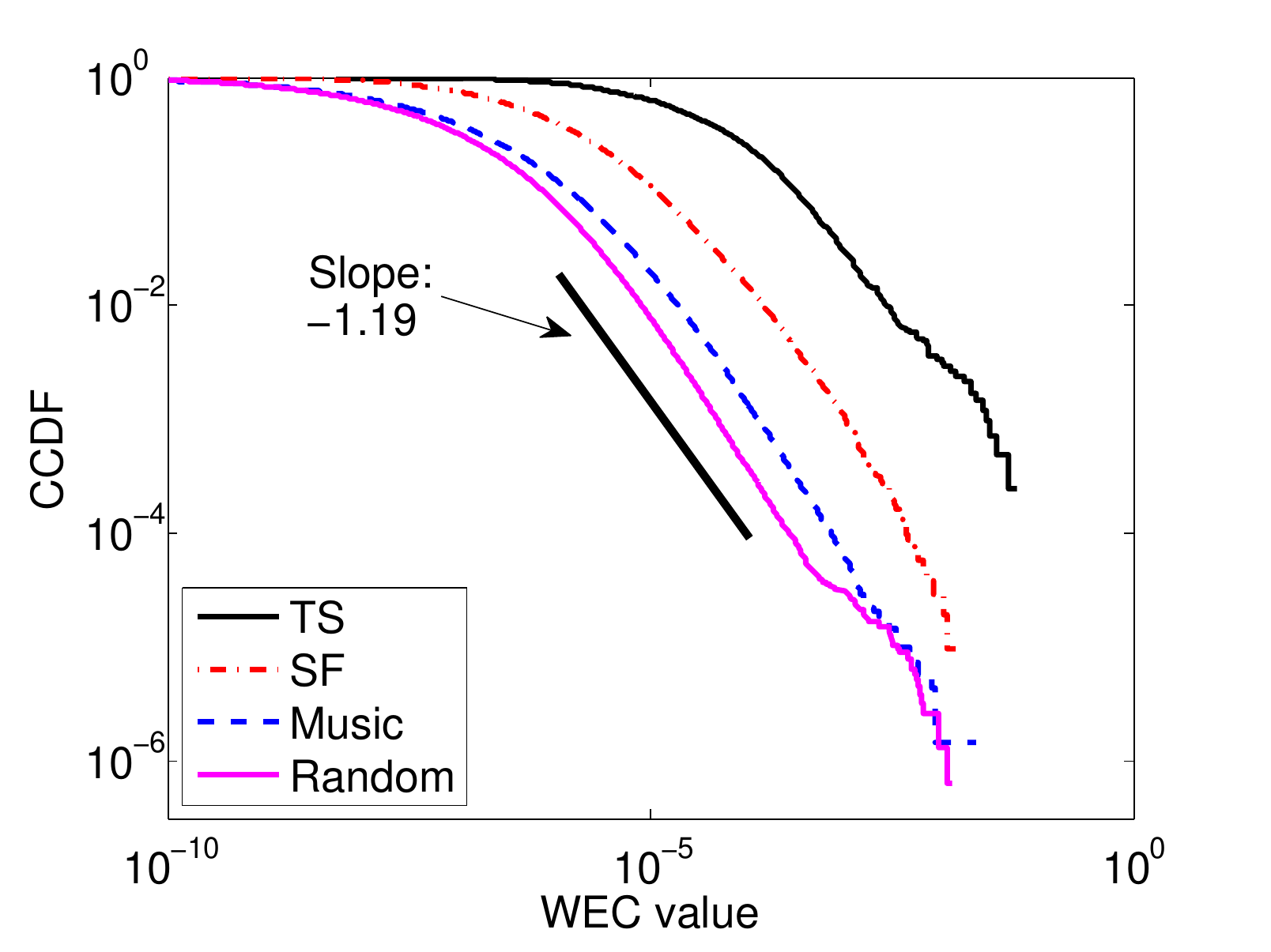}
    }
    \caption{The distribution of WEC values.}
    \label{fig:dist}
    \vspace{-.2in}
\end{figure}

%

\begin{figure*}[t]
\centering
 \subfloat[\texttt{SF} \label{fig:rwec-1-1}]{%
      \includegraphics[width=0.24\textwidth]{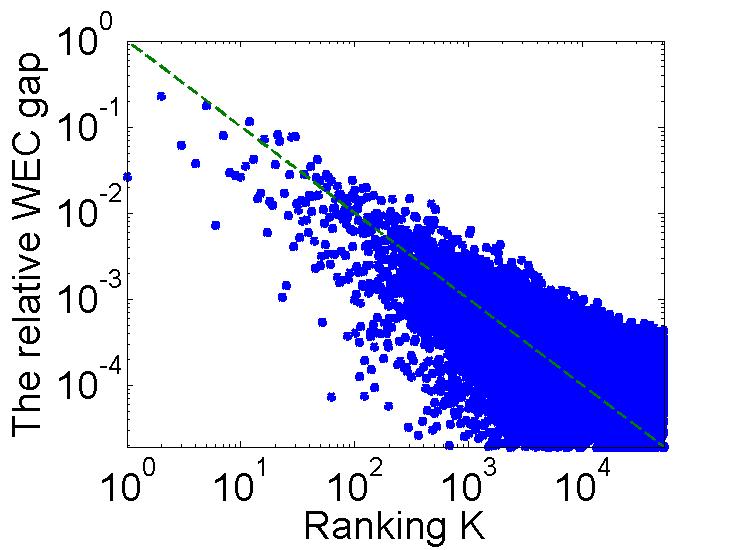}
    }
    \hfill
 \subfloat[\texttt{TS} \label{fig:rwec-2-1}]{%
      \includegraphics[width=0.24\textwidth]{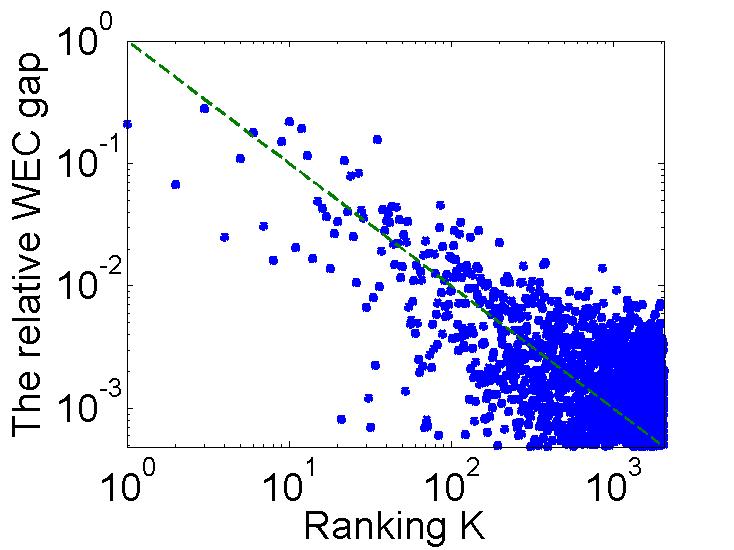}
    }
    \hfill
 \subfloat[\texttt{Music} \label{fig:rwec-3-1}]{%
      \includegraphics[width=0.24\textwidth]{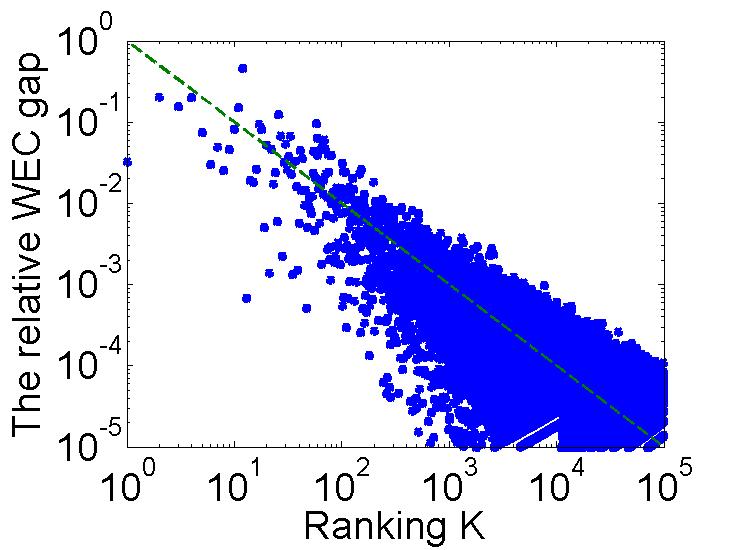}
    }
    \hfill
 \subfloat[\texttt{Random} \label{fig:rwec-4-1}]{%
      \includegraphics[width=0.24\textwidth]{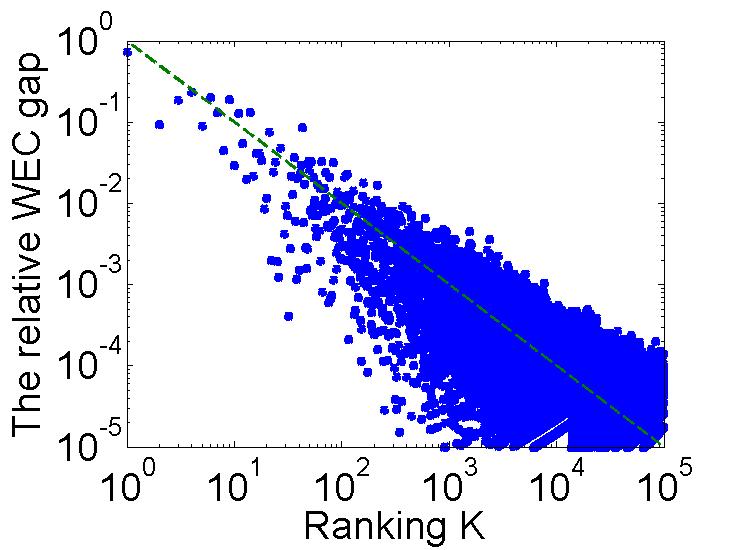}
    }
    \caption{Relative WEC gap $\Delta'_k$.}
    \label{fig:rwec}
    \vspace{-.2in}
\end{figure*}

\begin{figure*}[t]
\centering
 \subfloat[Following \label{fig:inactive1}]{%
      \includegraphics[width=0.2\textwidth]{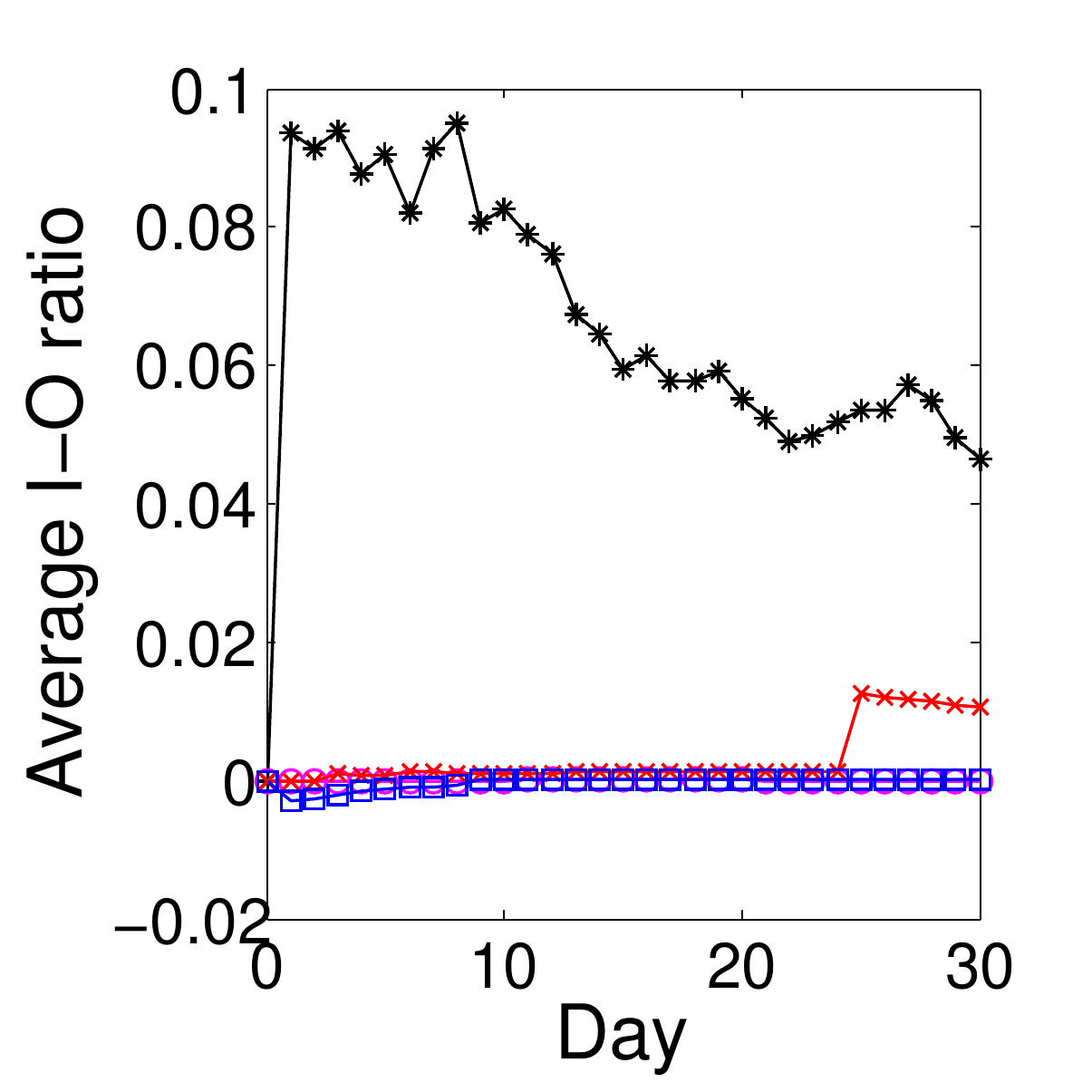}
    }
 \subfloat[Tweeting \label{fig:inactive2}]{%
      \includegraphics[width=0.2\textwidth]{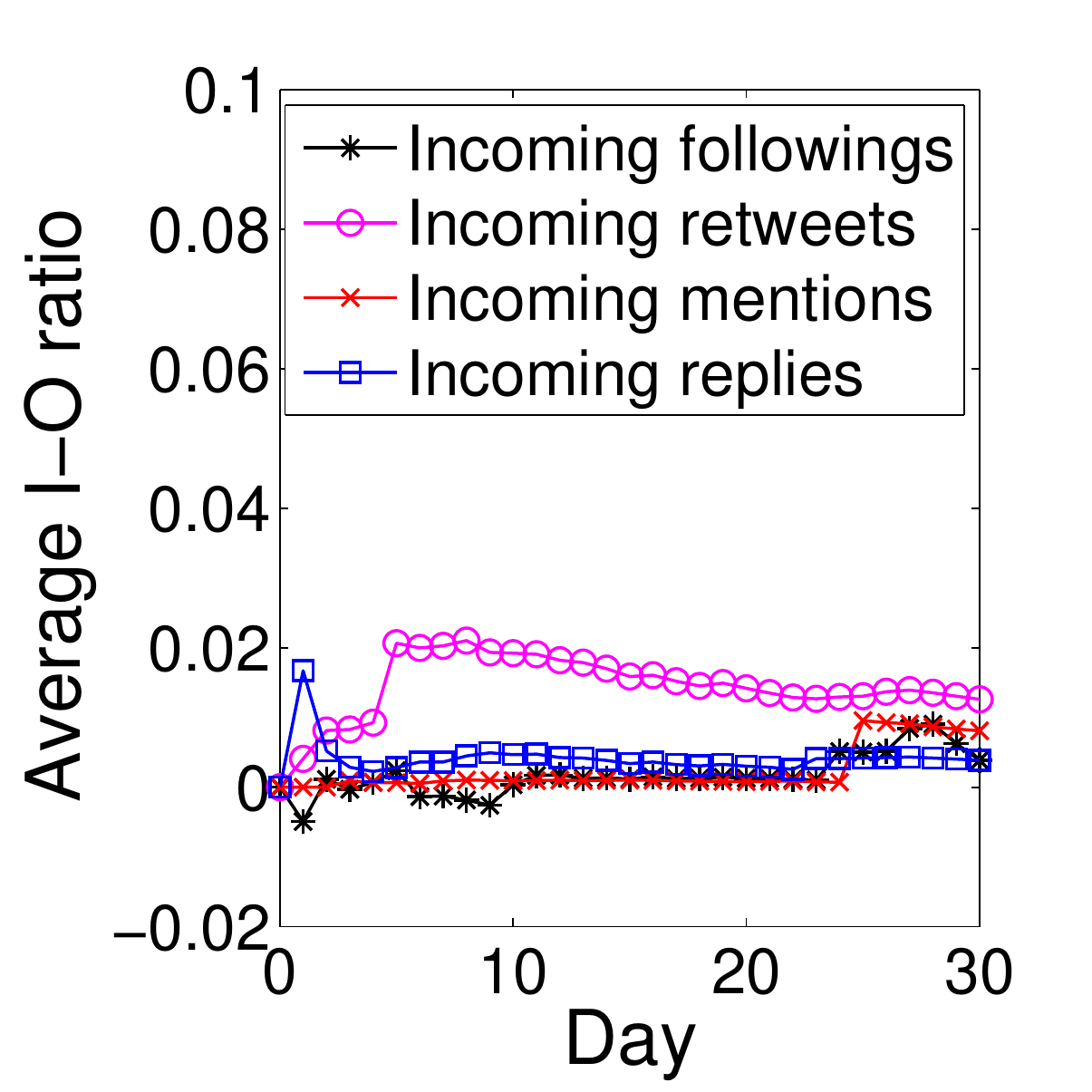}
    }
 \subfloat[Retweeting \label{fig:inactive3}]{%
      \includegraphics[width=0.2\textwidth]{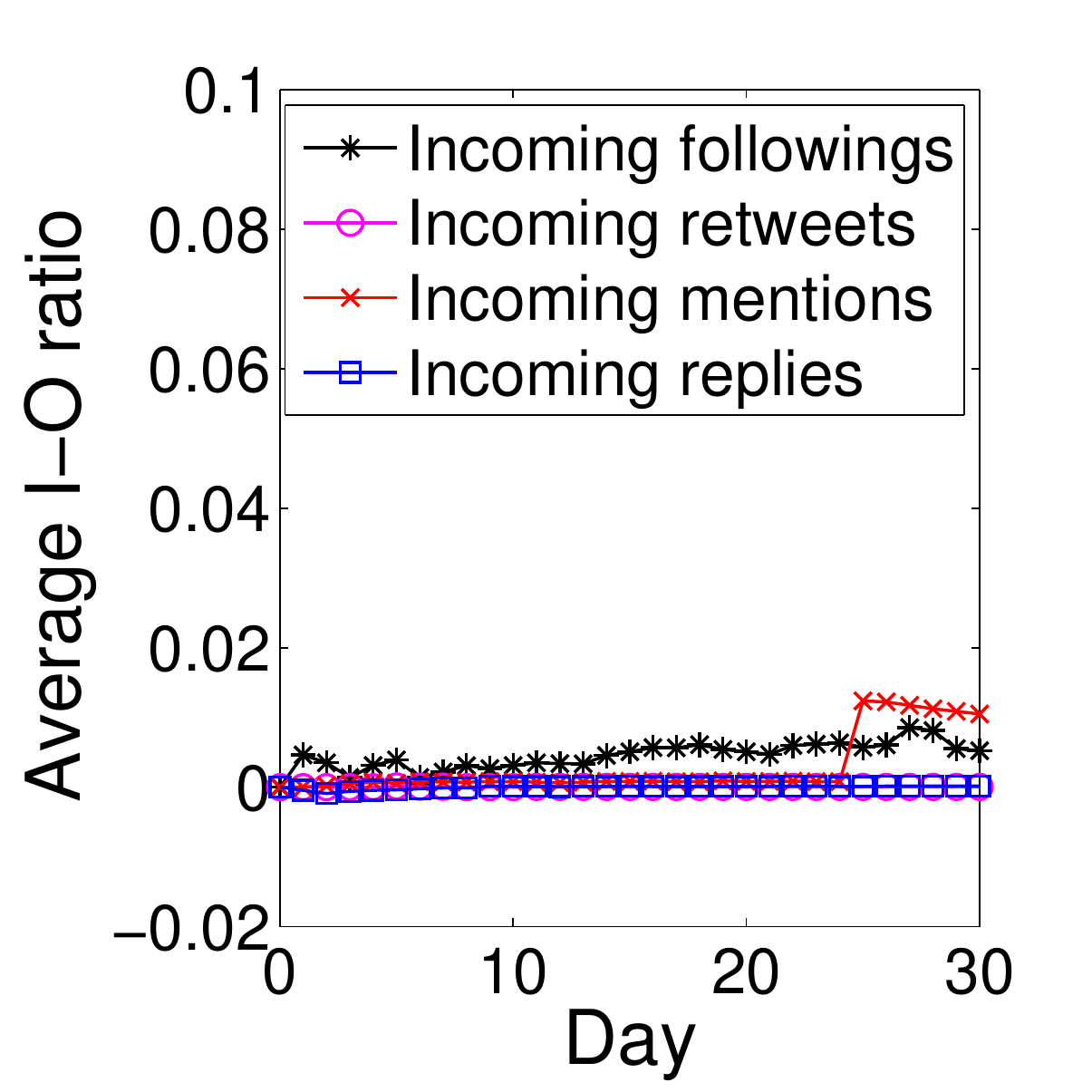}
    }
 \subfloat[Mentioning \label{fig:inactive4}]{%
      \includegraphics[width=0.2\textwidth]{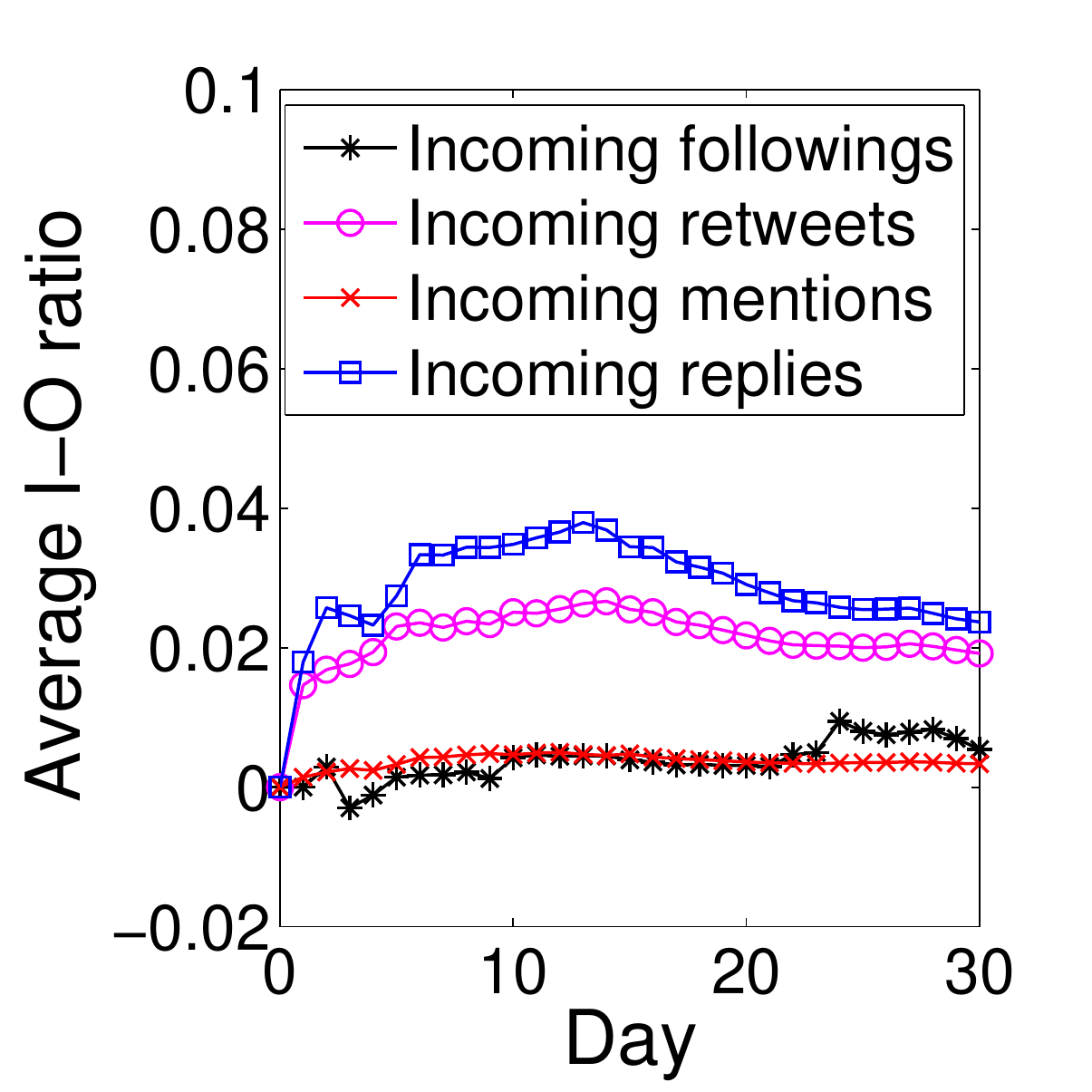}
    }
 \subfloat[Replying \label{fig:inactive5}]{%
      \includegraphics[width=0.2\textwidth]{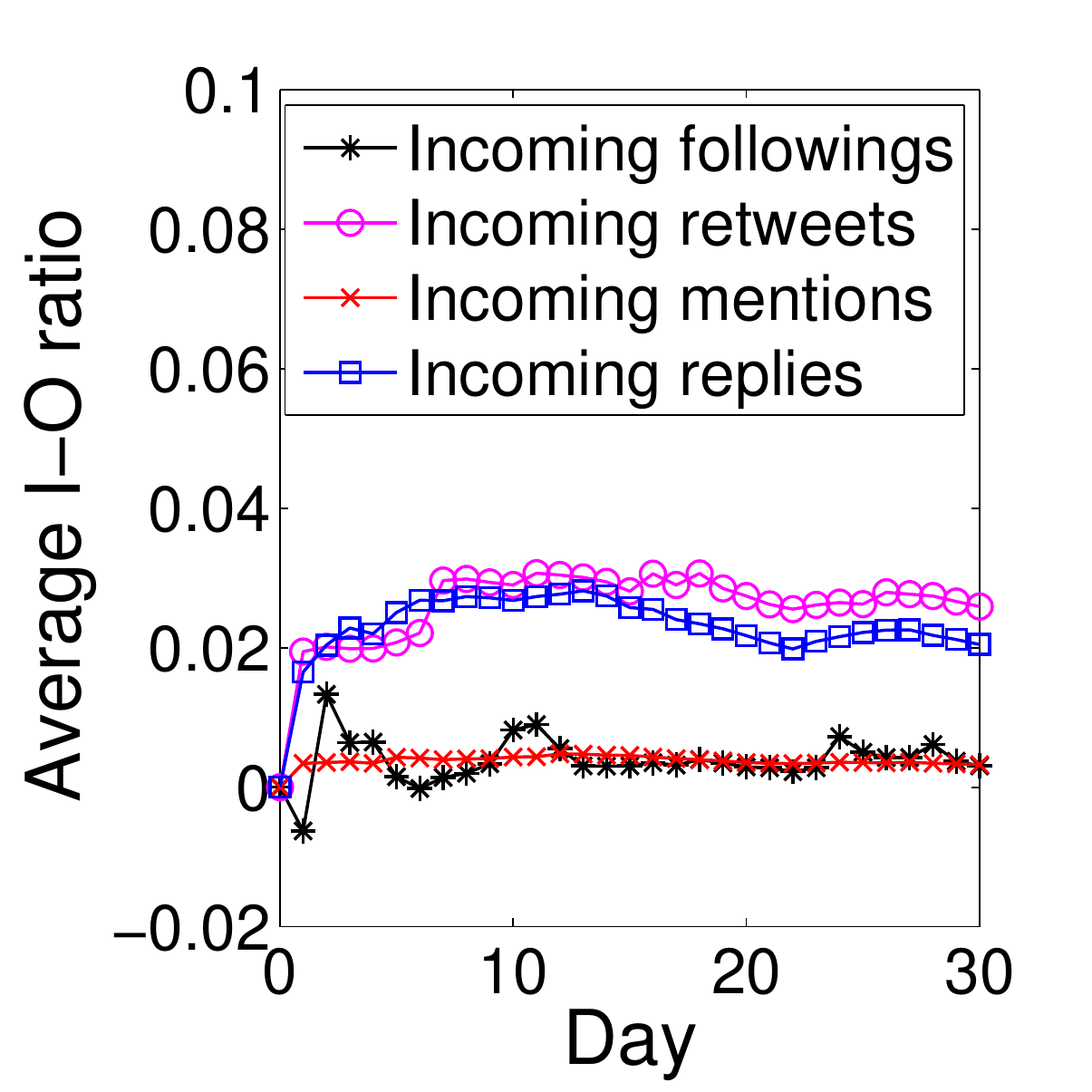}
    }
    \caption{Incoming-outgoing ratios for sybil groups, where the same legend is used in all the figures.}
    \label{fig:imbalance}
     \vspace{-.2in}
\end{figure*}

\subsection{Feasibility Studies} \label{sec:feasibility}

\subsubsection{WEC Value Characteristics}

TrueTop bases its early termination of iterative credit distribution on two assumptions. First, the WEC values of non-sybil nodes follow a power-law distribution. Second, the relative WEC gap $\Delta'_k$ decreases as $k$ increases. Now we verify these two assumptions.

Fig.~\ref{fig:dist} shows the log-log CCDF of the WEC values. We can see that all the CCDF curves are close to straight lines with the slopes from -2 to -1 for the WEC values larger than $10^{-6}$. Since a power-law distribution with PDF $p(x)=(\gamma -1) x^{-\gamma}$ has a CCDF $\bar{F}(x)=x^{1-\gamma}$, the WEC values of each interaction graph follow a power-law distribution with parameter $\gamma$ from 2 to 3.

Fig. \ref{fig:rwec} shows the log-log scale of $\Delta'_k$ as a function of $k$, where the results are shown up to $k=10^5$ due to space constraints. We computed the WEC values by
using $\nu=10^{-4}$ as the error tolerance threshold of power iterations, which led to about 1,000 iterations. $\Delta'_k$ obviously decreases with an approximate slope of -1 in the log-log scale, which coincides well with the analysis in Lemma~\ref{th:gap}.

\subsubsection{Interaction Analysis} \label{sec:interactionAnalysis}

Since there is no benchmark for the real-world sybils on Twitter, we designed an experiment to estimate the total edge weight from the non-sybil region to the sybil region in order to verify that it is relatively very small. To catch the growing intelligence of Twitter sybils, we adopted the behavior of the emerging social bots \cite{MessiYou13,ZhangOn13,FerraRis14}. Our experiment run as follows. We first purchased 1000 Twitter accounts, then divided them to mimic legitimate activities as in \cite{MessiYou13, ZhangOn13}, and finally investigated how many legitimate users will follow or interact with them. Specifically, we divided these 1000 accounts into five groups of equal size, each corresponding to a unique activity among following, tweeting, retweeting, mentioning, and replying. We ran the experiment for 30 days. In each day, we let each sybil user in each group initiate 10 activities corresponding to that group. For example, each sybil user in the Following group followed 10 randomly-chosen new users in each dataset every day. Except the sybil users in the Tweeting group, the sybil users in all the other groups initiated the corresponding activities only towards randomly chosen new users in each dataset. We also recorded the total followings/mentions/retweets/replies every sybil group received each day. In addition, we chose the \texttt{Random}, \texttt{SF}, and \texttt{Music} datasets as the target datasets in the first 14, middle 8, and last 8 days, respectively.

Fig.~\ref{fig:imbalance} shows the incoming-outgoing (I-O) ratios of each sybil group, which is defined as the number of total followings/mentions/retweets/replies each sybil group received every day over the total number of interactions initialized from the sybil group in the same day (i.e., 2,000). We have two observations. First, non-sybil users are very careful about whom to interact with and rarely interact with sybil users. Second, sybil users can get a non-trivial number of non-sybil followers. We manually found that most non-sybil followers are normal users out of reciprocity, social capitalists, or even spam accounts not suspended by Twitter, and this observation is in line with prior results in \cite{GhoshUnd12,YangAna12}. So incoming followings are less trustworthy for evaluating user influence than incoming replies, mentions, and retweets.

To compute the I-O ratios of the sybil and non-sybil communities, we randomly chose 30 groups of 200 users from each of \texttt{Random}, \texttt{SF}, and \texttt{Music} datasets. We then recorded the incoming and outgoing interactions of each non-sybil group every day in the same experimental period. The I-O ratio for each sybil or non-sybil group is redefined as the total incoming edge weight over the total outgoing edge weight. Table~\ref{tlb:imbalance} compares the average I-O ratios of the sybil and non-sybil groups for both sum-based and entropy-based edge weights. As we can see, non-sybil communities always have much higher I-O ratios (i.e., much more balanced incoming and outgoing interactions) than sybil communities. Moreover, the entropy-based weight model yields lower and higher I-O ratios than the sum-based weight model for the sybil and non-sybil communities, respectively. We thus expect that the entropy-based weight model can lead to better sybil resilience than the sum-based model (as shown in Table~\ref{tlb:option}).

\begin{table}[t]
    \caption{The comparison of incoming-outgoing ratios between sybil and non-sybil communities under sum-based and entropy-based interaction graphs.}
    \centering
    \begin{tabular}{|l|l|r|r|r|}
       \hline
       Graph Model & Community &\texttt{SF} & \texttt{Random} & \texttt{Music}\\
       \hline
       \multirow{2}{*}{Sum} & Non-sybil & 0.89 &  1.04 & 0.70 \\
       \cline{2-5}
                                 & Sybil & 0.08 & 0.08 & 0.08 \\
       \hline
       \multirow{2}{*}{Entropy} & Non-sybil & 1.54 & 1.15 & 0.60 \\
       \cline{2-5}
                                  & Sybil & 0.04 & 0.07 & 0.05 \\
        \hline
    \end{tabular}
    \label{tlb:size} \label{tlb:imbalance}

     \vspace{-.2in}
\end{table}

\subsection{Accuracy and Sybil Resilience Studies}\label{sec:Performance}

\subsubsection{Evaluation Methodologies}

Since large-scale real experiments on Twitter inevitably violate the Twitter ToS, we resort to synthetic simulations to evaluate the accuracy and sybil resilience of TrueTop. We used all the four datasets and obtained quite consistent results. Below we show the evaluation results for the \texttt{SF} dataset only due to space constraints.

We modelled the strength of sybil attacks on Twitter by a parameter $\alpha$, which refers to the ratio of the total edge weight in the non-sybil region over that from the non-sybil region to the sybil region. The default value of $\alpha$, denoted by $\alpha^*$, is obtained from our datasets as follows. Assume that the network is composed of a non-sybil region with $n_1$ twitterers and a sybil region with $n_2$ twitterers. According to our experiments, we found that about 0.98\textperthousand\hspace{0.05cm} of the users in the \texttt{SF} dataset have been suspended, so we set $n_1 = 1000n_2$. Moreover, assume that each non-sybil user initiate one interaction (i.e., retweeting, mentioning, or replying) to each of the other $n_1-1$ users, leading to $n_1(n_1-1)$ outgoing interactions. According to Table~\ref{tlb:imbalance}, the average I-O ratio of the non-sybil community for the sum-based interaction network is $(0.89+1.04+0.7)/3\approx 0.88$. Therefore, the $n_1$ non-sybil users can receive about $1.88 n_1(n_1-1)\approx 1.88n_1^2$ incoming and outgoing interactions. Similarly, the sybil users issue totally $n_2n_1$ interactions to the non-sybil region and receive about $0.08 n_2n_1$ interactions from non-sybil users. We thus have the following approximation \begin{equation}
\alpha^* = \frac{0.08 n_2n_1}{1.88n_1^2} = \approx 4.2 * 10^{-5}.
\end{equation}

We used the following method to simulate the sybil region, which has been adopted in \cite{YuSyb06, CaoAid12}. Given the interaction graph constructed from the \texttt{SF} dataset, we can expect that the majority of the 104,000 users there are non-sybil users, but we cannot tell which users are sybil or non-sybil users. So we manually attached to the original interaction graph a sybil region which is a complete digraph of 500 sybil users and ran TrueTop over this augmented interaction graph. We assume the worst-case scenario in which the attacker aims to retain all the credits flowing into the sybil region, so there is no interaction from the sybil region to the non-sybil region. We then added $w_g$ random links of weight one from the non-sybil region to the sybil region, which is equivalent to assuming that there are $w_g$ accidental one-time interactions from non-sybil users to sybil users. $w_g$ varied from 10 to 200 in our experiments. Since the total edge weight of the original interaction graph is about $10^6$, we effectively simulated the parameter $\alpha$ from $10^{-5}$ to $2\times 10^{-4}$. To simplify the presentation, we equate $w_g$ with $\alpha$ and call it the attack strength as well hereafter.

We considered three strategies for the attacker to add the $w_g$ links. In the \emph{random attack}, the attacker randomly selects $w_g$ users in the non-sybil region and adds a link of weight one from each to a randomly chosen user in the sybil region. In the \emph{community attack}, the attacker performs a breadth-first search from a random user in the non-sybil region until $w_g$ users are found, and it adds a link from each discovered user to a random user in the sybil region. In the \emph{seed attack}, we fixed 10 seed users in the non-sybil region and assumed that the attacker knows all of them. The attacker performed a breadth-first search from the 10 seed users and randomly chose $w_g$ users closest to any of the 10 seed users. It finally adds a link of weight one from each of them to a random user in the sybil region. Obviously, the seed attack corresponds to the strongest attack. We conducted 50 experiments for each attack and report the average result below. In addition, we chose 100 verified users as seed users in all simulations.

Now we introduce some metrics to measure the accuracy and sybil resilience. Recall that $\mathcal{U}_K$, $\mathcal{U}_K^\ast$, and $\tilde{\mathcal{U}}$ denote the TrueTop output, the true top-$K$ influential users in the non-sybil region, and all the sybil users, respectively. We obtained $\mathcal{U}_K^\ast$ by running power iteration over the non-sybil region only with the error tolerance $\nu=10^{-8}$. We measure the accuracy of TrueTop by comparing $\mathcal{U}_K$ and $\mathcal{U}_K^\ast$ via the following two types of errors.
\begin{itemize}
\item Type-I error: $d(K)/K$, where $d(K)$ is the distance between $\mathcal{U}_K$ and $\mathcal{U}_K^\ast$ and computed according to Eq.~(\ref{eq:distance}). The metric measures the average rank offset of $\mathcal{U}_K^\ast$ from  $\mathcal{U}_K$.
\item Type-II error: $(K - |\mathcal{U}_K^\ast \cap \mathcal{U}_K|)$. This metric measures how many true top-$K$ users are missed by TrueTop.
\end{itemize}
The sybil resilience of TrueTop is inversely proportional to $\#_{\textsf{sybil}}=|\tilde{\mathcal{U}}\cap \mathcal{U}_K|$. After iterative credit distribution in TrueTop terminates, assumes that totally $C$ credits are retained in the sybil region. Let $C_1,\dots,C_K$ denote the credits of the top-$K$ influential users in the non-sybil region in a non-decreasing order. Also assume that the attacker tries to maximize $\#_{\textsf{sybil}}$ by arbitrarily manipulating the topology of the sybil region such that the $C$ credits can flow into a few sybil users. We can derive $\#_{\textsf{sybil}}$ as follows:

\begin{equation*}
\#_{\textsf{sybil}} = \left\{
    \begin{array}{rl}
        0 & \text{if } C < C_K, \\
        \underset{1\leq x \leq K}{ \text{argmax}} \quad \quad C \geq x C_{K + 1 - x} & \text{else}.
    \end{array} \right.
\end{equation*}

\subsubsection{Basic Results}

\begin{figure}[t]
\centering
 \subfloat[ Random attack\label{fig:sybil-str-1}]{%
      \includegraphics[width=0.25\textwidth]{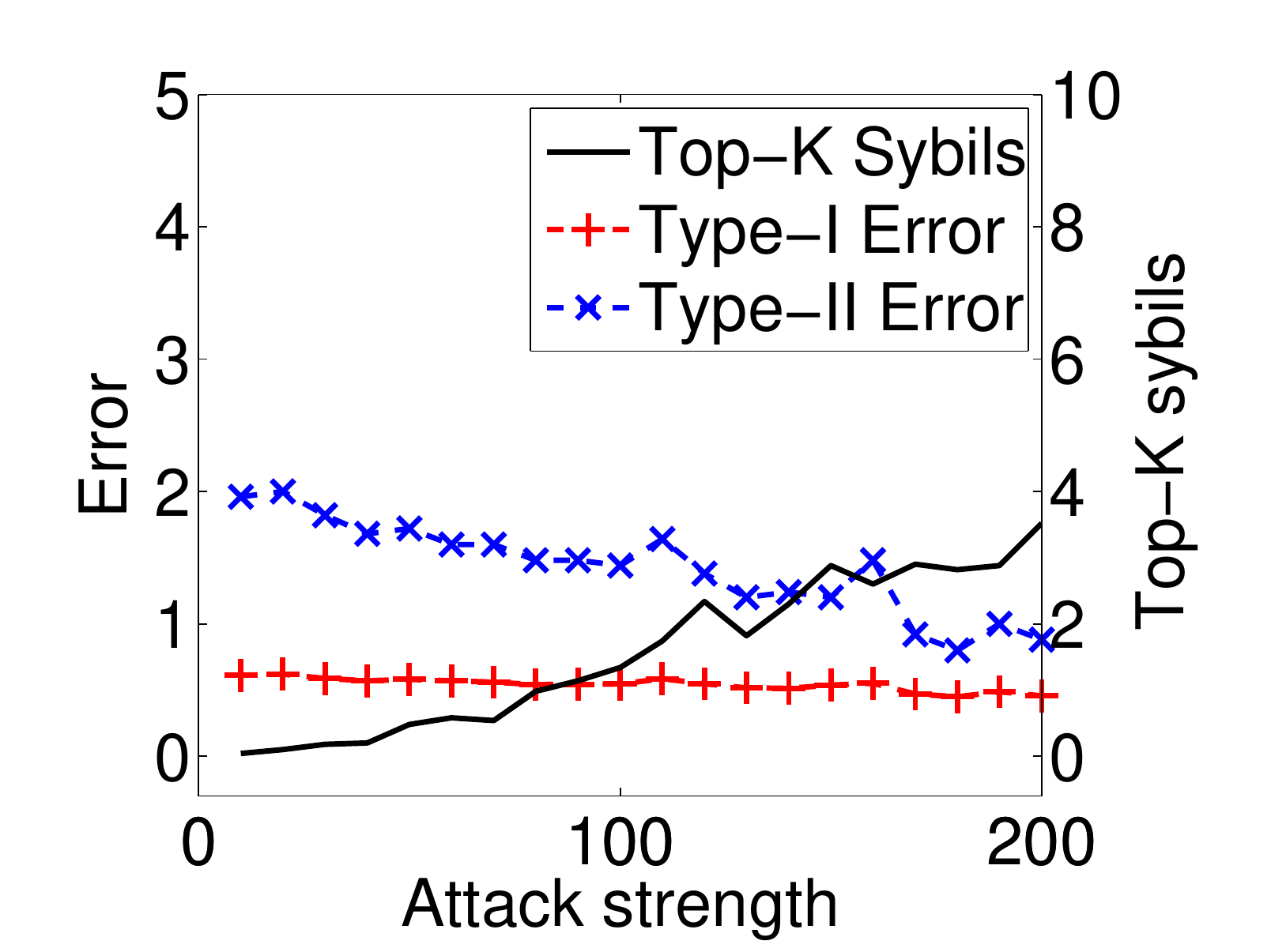}
    }
 \subfloat[Community attack \label{fig:sybil-str-2}]{%
      \includegraphics[width=0.25\textwidth]{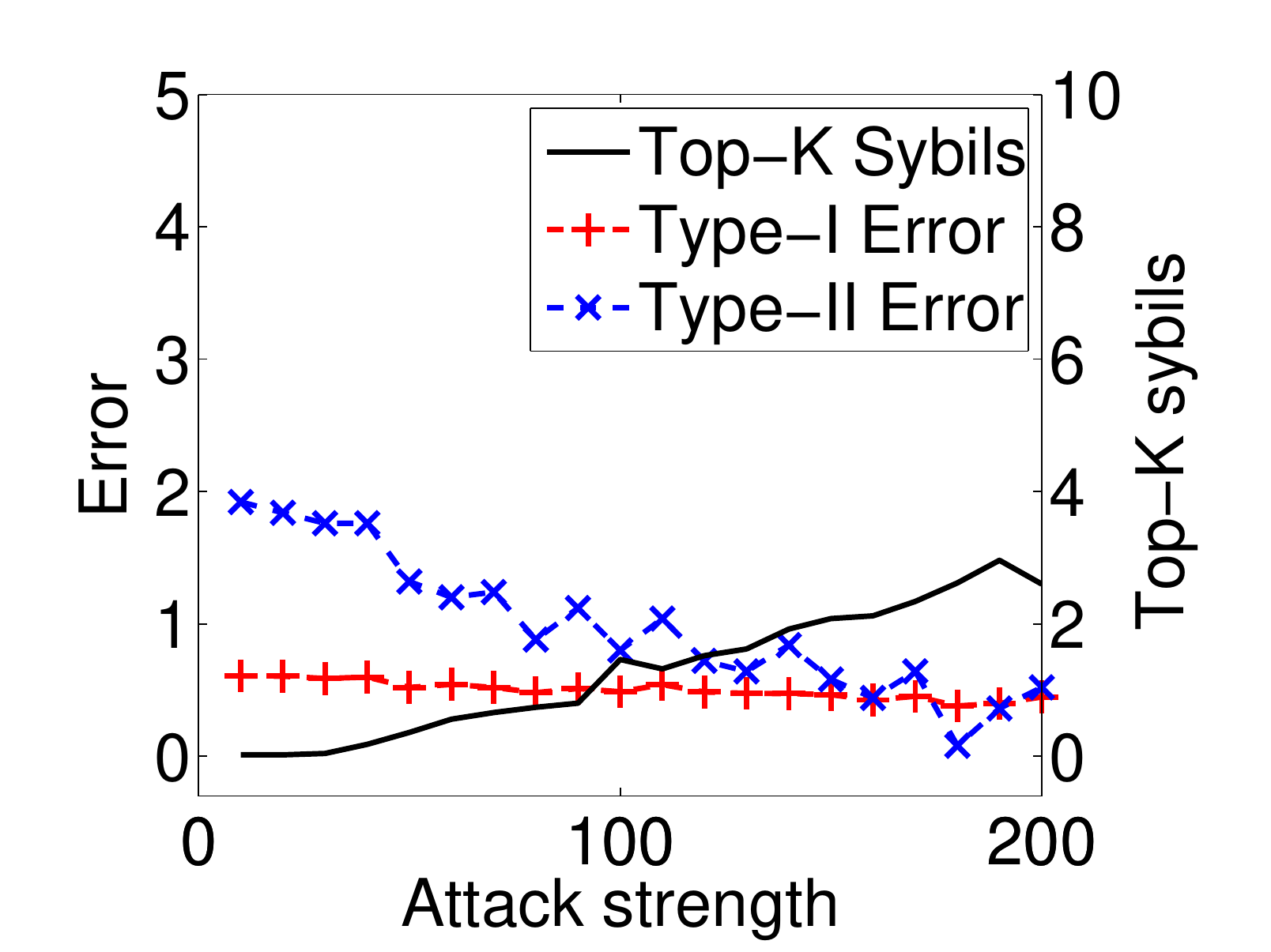}
    }
    \caption{TrueTop performance under different attack strengths}
    \label{fig:sybil-str}
     \vspace{-.2in}
\end{figure}

\begin{figure}[t]
\centering
 \subfloat[ Random attack\label{fig:sybil-k-1}]{%
      \includegraphics[width=0.25\textwidth]{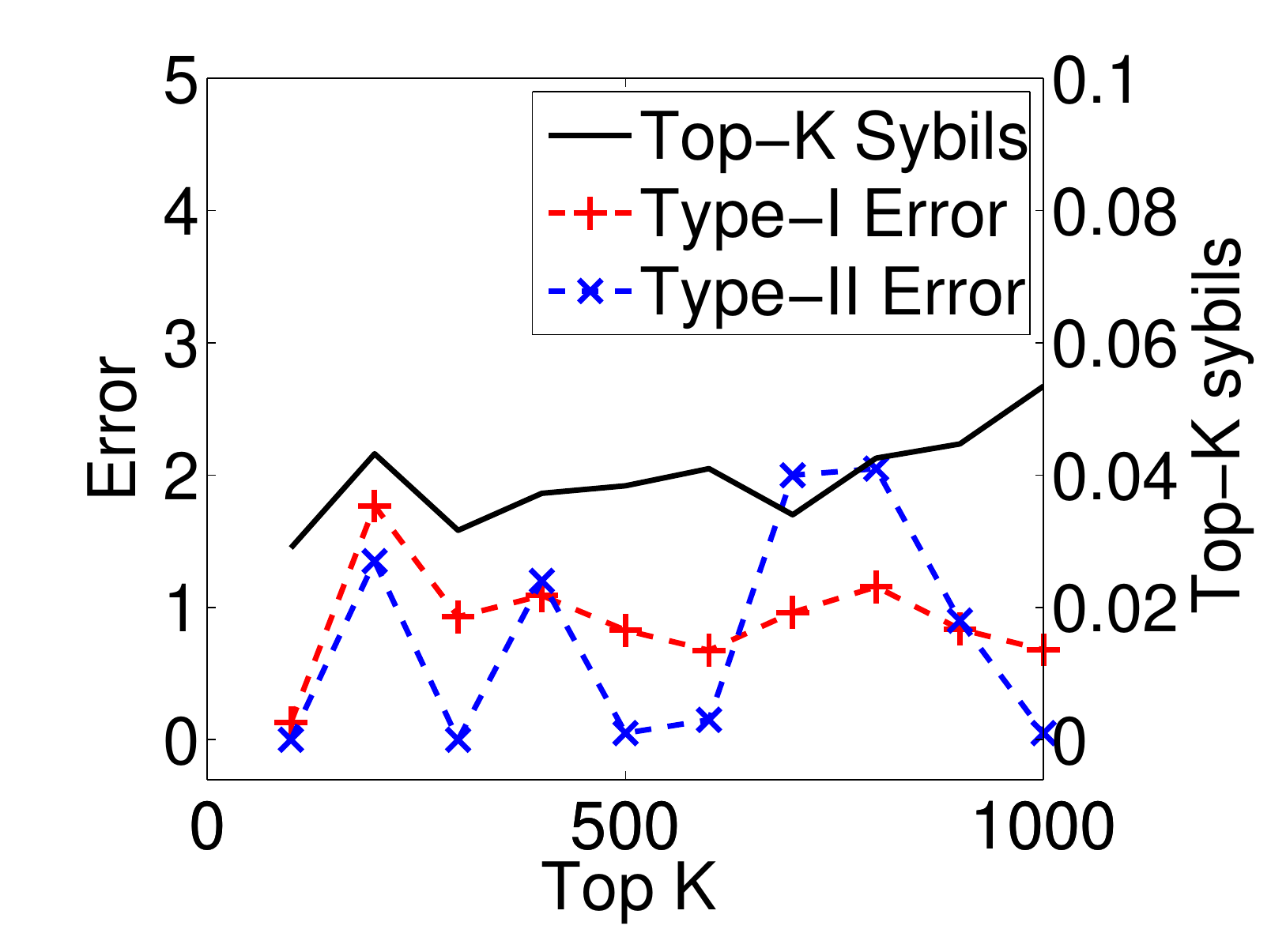}
    }
 \subfloat[Community attack \label{fig:sybil-k-2}]{%
      \includegraphics[width=0.25\textwidth]{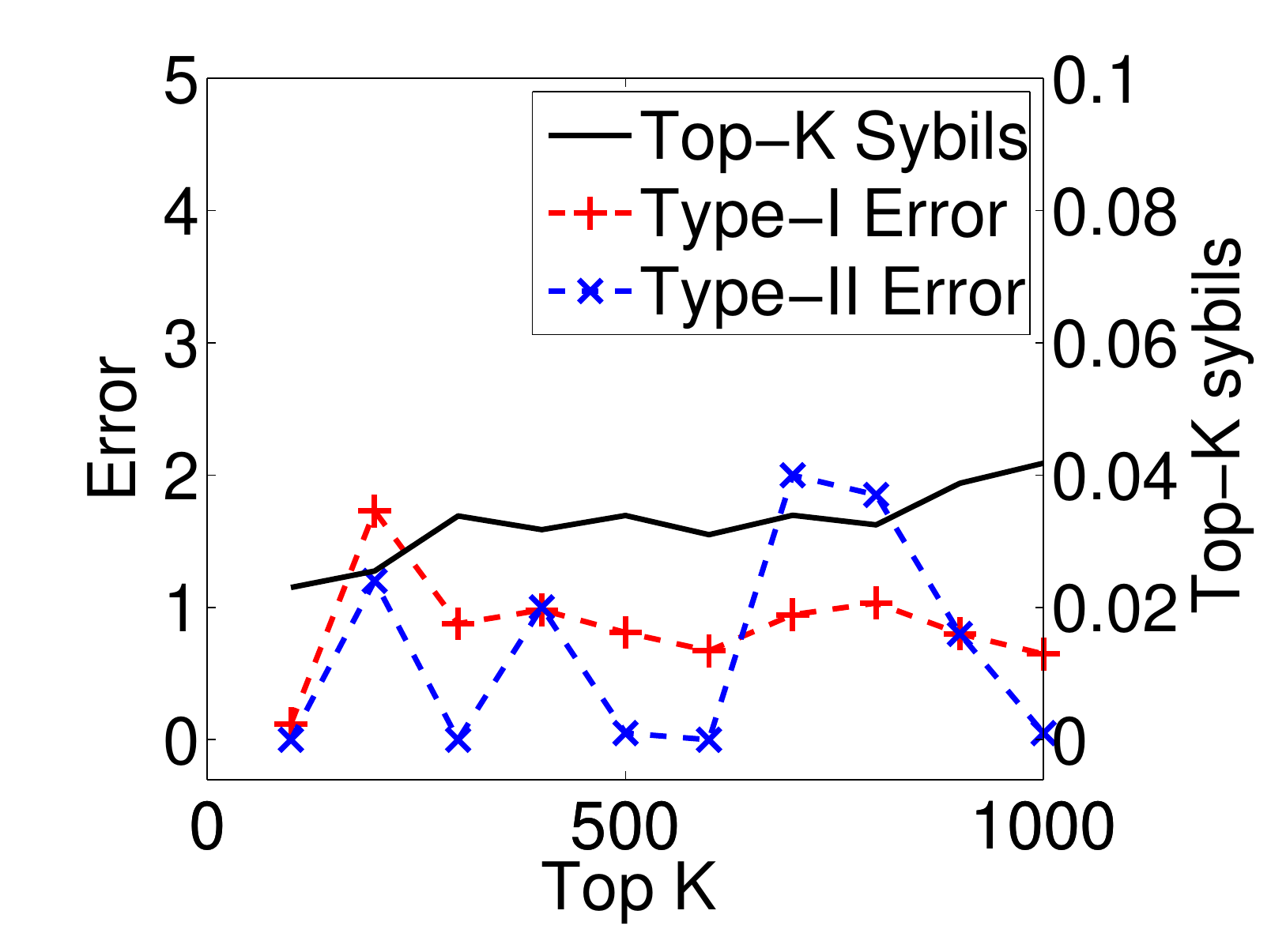}
    }
    \caption{TrueTop performance for different $K$s.}
    \label{fig:sybil-k}
     \vspace{-.2in}
\end{figure}

\begin{figure}[t]
\centering
 \subfloat[ Random attack\label{fig:sybil-epsilon-1}]{%
      \includegraphics[width=0.25\textwidth]{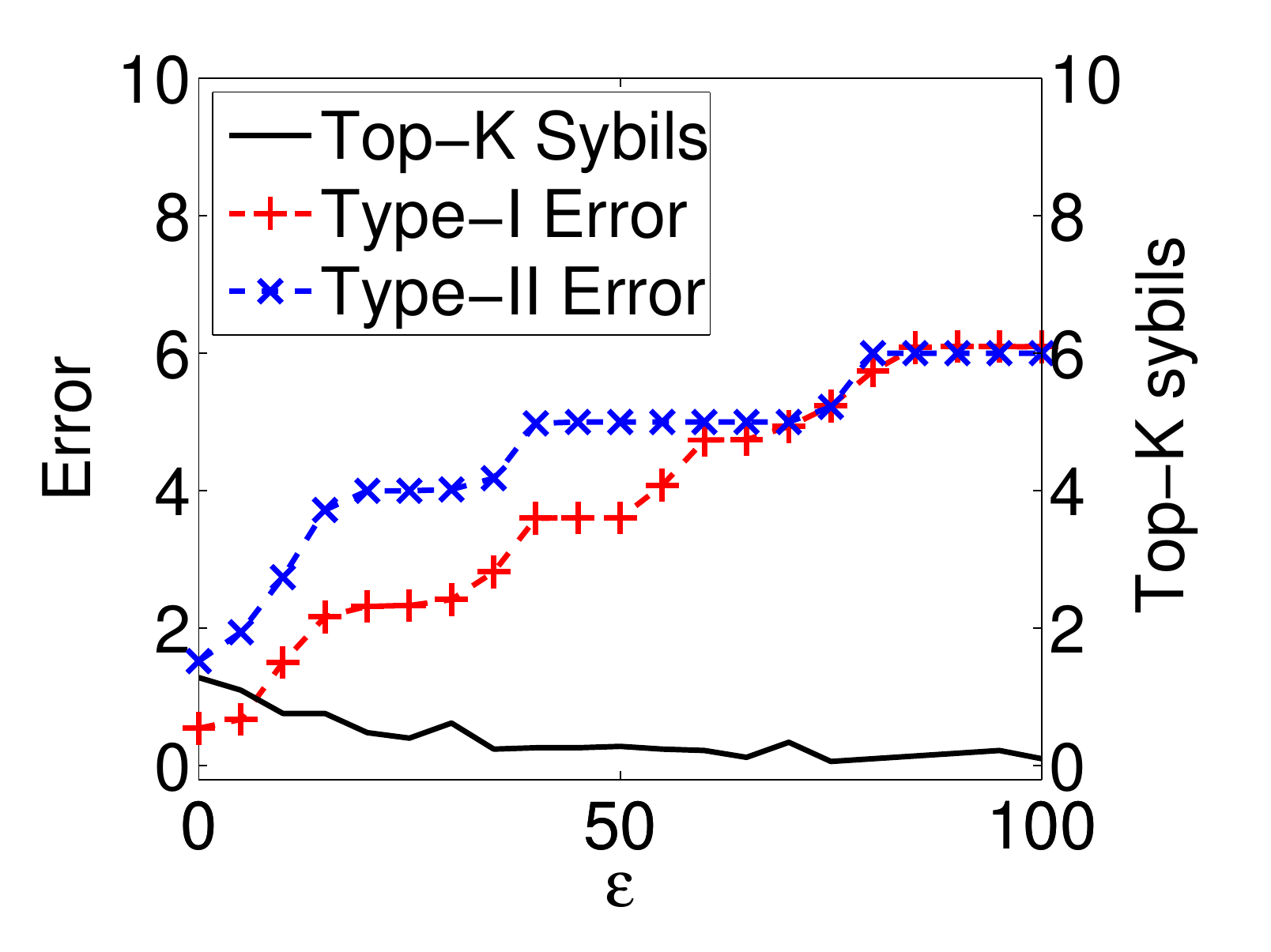}
    }
 \subfloat[Community attack \label{fig:sybil-epsilon-2}]{%
      \includegraphics[width=0.25\textwidth]{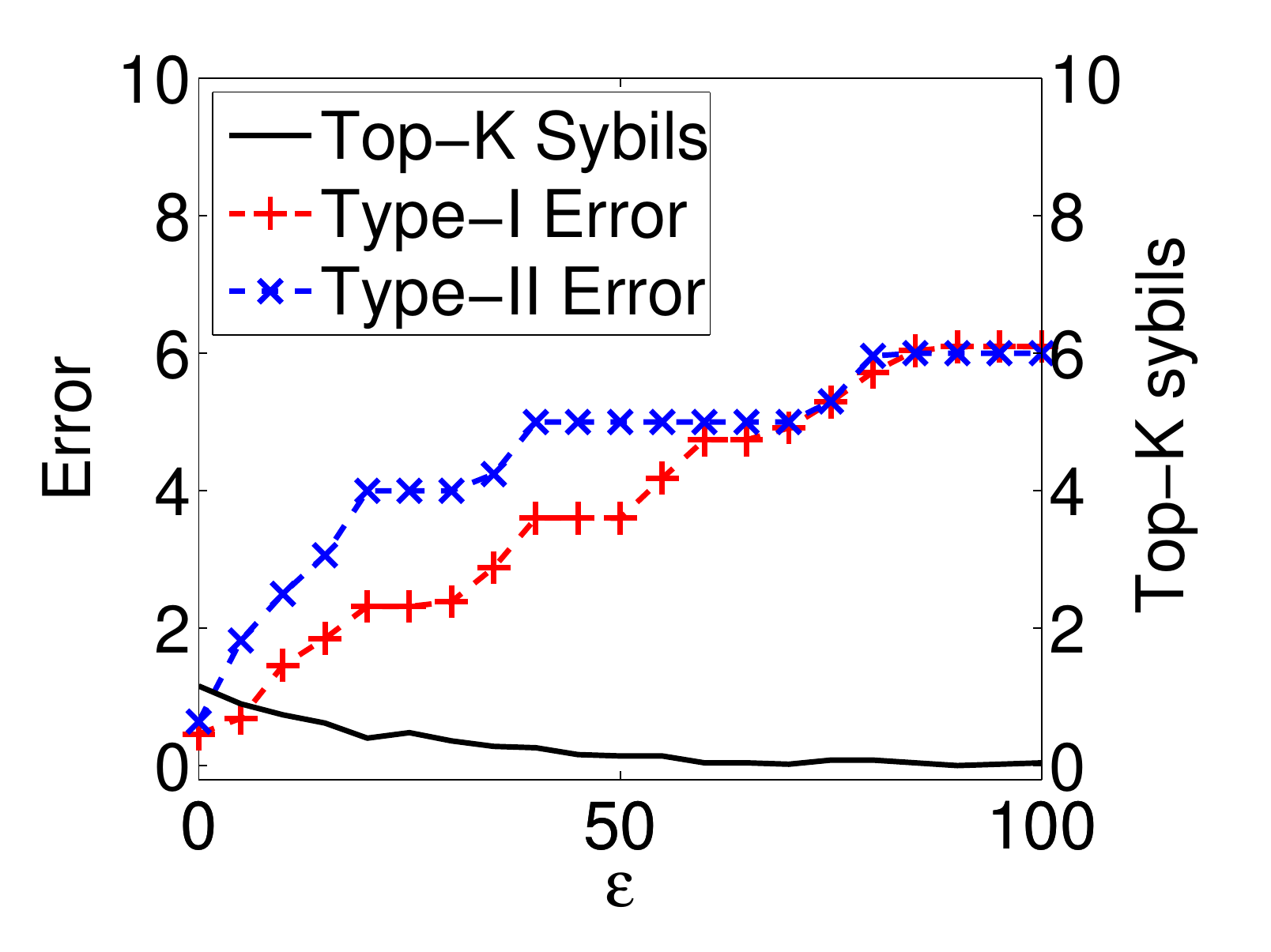}
    }
    \caption{TrueTop performance under different $\epsilon$s.}
    \label{fig:sybil-epsilon}
    \vspace{-.2in}
\end{figure}

\begin{figure}[t]
\centering
 \subfloat[Sum-based \label{fig:sybil-weight-1}]{%
      \includegraphics[width=0.25\textwidth]{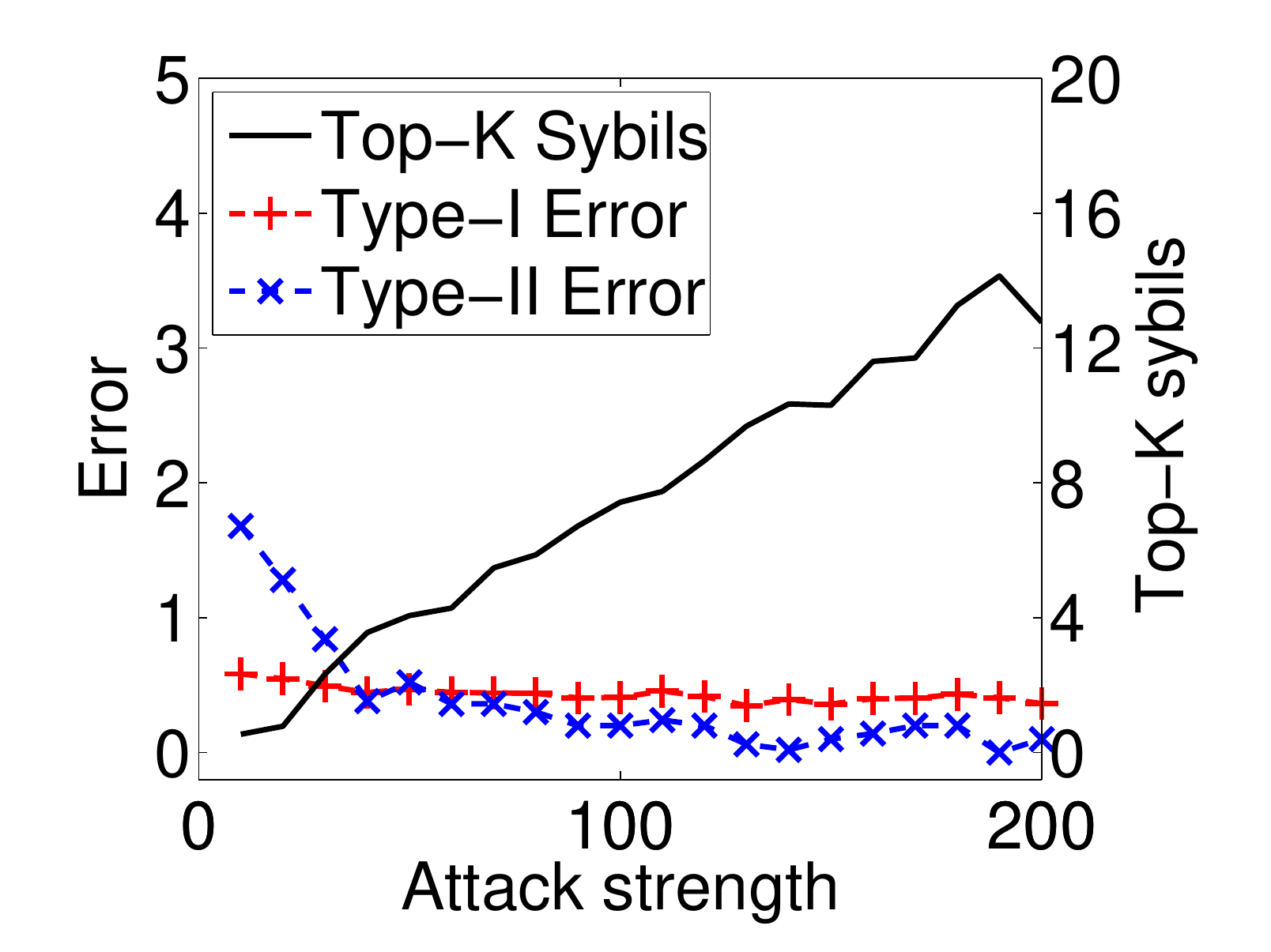}
    }
 \subfloat[Entropy-based \label{fig:sybil-weight-2}]{%
      \includegraphics[width=0.25\textwidth]{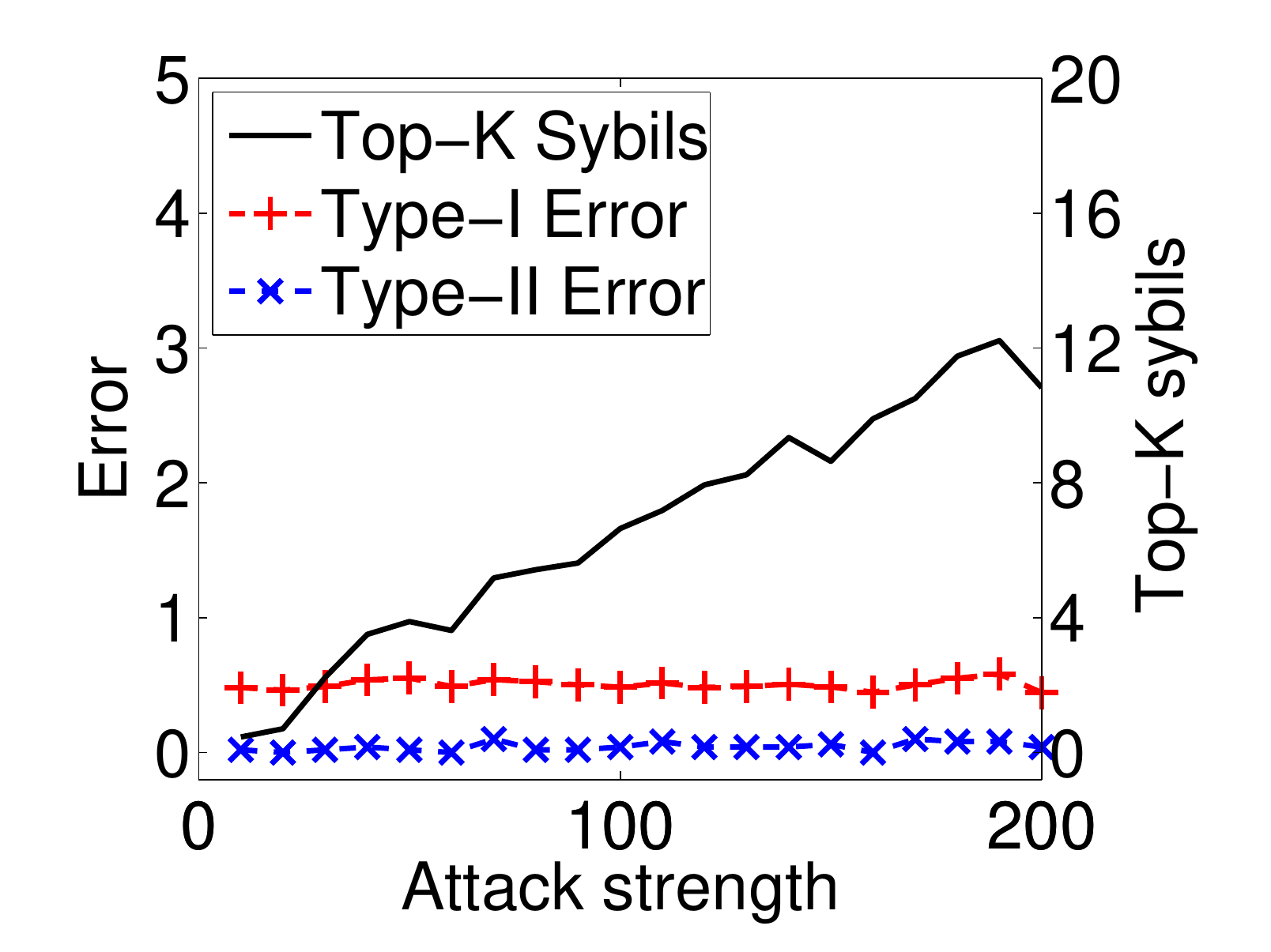}
    }
    \caption{Impact of seed attacks with different weight models.}
    \label{fig:sybil-weight}
     \vspace{-.2in}
    \end{figure}


\begin{table*}[t]
    \centering
    \caption{The impact of different design options on TrueTop performance.} 
\begin{tabular}{|c|c|c|c|c|c|c|c|c|c|}
       \hline
        & \multicolumn{3}{|c|}{Random attack} & \multicolumn{3}{|c|}{Community attack} & \multicolumn{3}{|c|}{Seed attack}\\
       \hline
        Seed selection: \newline basic vs. rwec & 0.11 &  0.232  &  -0.017 & -0.03  &  -0.19 & -0.192 & 0.11  &   0.19  &  0.316  \\
        Edge weights: \newline sum vs. entropy & 0.07  &  -0.002  &  0.226 & 0.08  &  0.00  &  0.572 & -0.071  &  0.327  &  0.871  \\
        $\#$ of seeds: \newline 10 vs. 100 & 4.26  &  0.099  &  0.122 & 2.89  &  0.121  &  0.078 & 3.66  &  0.136  &  2.8  \\
        \hline
        & Type-I  &  Type-II  &  $\#_{\textsf{sybil}}$ & Type-I &  Type-II  &  $\#_{\textsf{sybil}}$ & Type-I &  Type-II  &  $\#_{\textsf{sybil}}$  \\
        \hline
    \end{tabular}
    \label{tlb:size} \label{tlb:option}
     \vspace{-.2in}
\end{table*}

Fig.~\ref{fig:sybil-str} shows the performance of TrueTop under different attack strengths in random and community attacks. In this experiment, we set $K=100$ and $\epsilon=0$. As the attack strength increases from 10 to 200, the type-I error is flat with less than one, and the type-II error is below two, both showing the high accuracy of TrueTop under different attack strengths. Moreover, the number of top-$100$ sybil users, i.e., $\#_{\textsf{sybil}}$, slowly increases as $w_g$ increases, which is as expected. $\#_{\textsf{sybil}}$, however, stays below four for both attacks. In addition, larger $w_g$ is likely to increase the number of iterations and thus make the top-$K$ list more accurate. So we can see that the type-II error overall decreases with increasing $w_g$.

Fig.~\ref{fig:sybil-k} shows the performance of TrueTop under different $K$s in random and community attacks. In this experiment, we set the $w_g=100$ and $\epsilon=0$. We also normalized $\#_{\textsf{sybil}}$ by $K$. Although $\#_{\textsf{sybil}}/K$ slowly increases with $K$ due to more iterations, it is always less than 6\%. In addition, both type-I and type-II errors are always less than two, indicating the high accuracy of TrueTop.

Fig.~\ref{fig:sybil-epsilon} shows the performance of TrueTop under different $\epsilon$s in random and community attacks. In this experiment, we set $w_g=100$ and $K=100$. As expected, the larger the error tolerance $\epsilon$, the larger both type-I and type-II errors. In contrast, $\#_{\textsf{sybil}}$ decreases with increasing $\epsilon$ due to fewer iterations towards credit distribution termination.

\begin{figure*}
\centering
\begin{minipage}{0.48\textwidth}
\centering
\subfloat[Under random attacks \label{fig:sybil-compare-1}]{\includegraphics[width=0.5\textwidth]{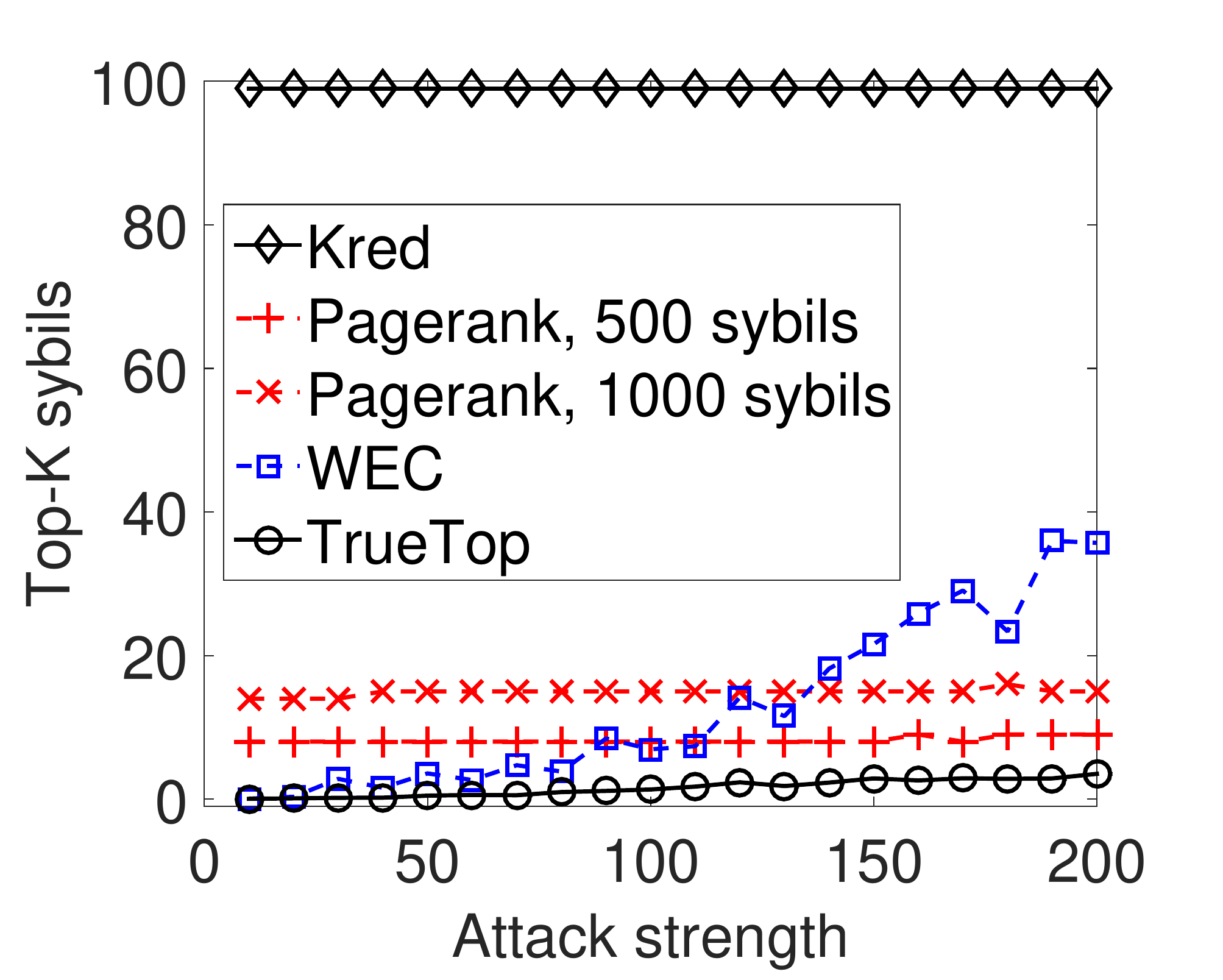}}\hfill
\subfloat[Under community attacks \label{fig:sybil-compare-2}]{\includegraphics[width=0.5\textwidth]{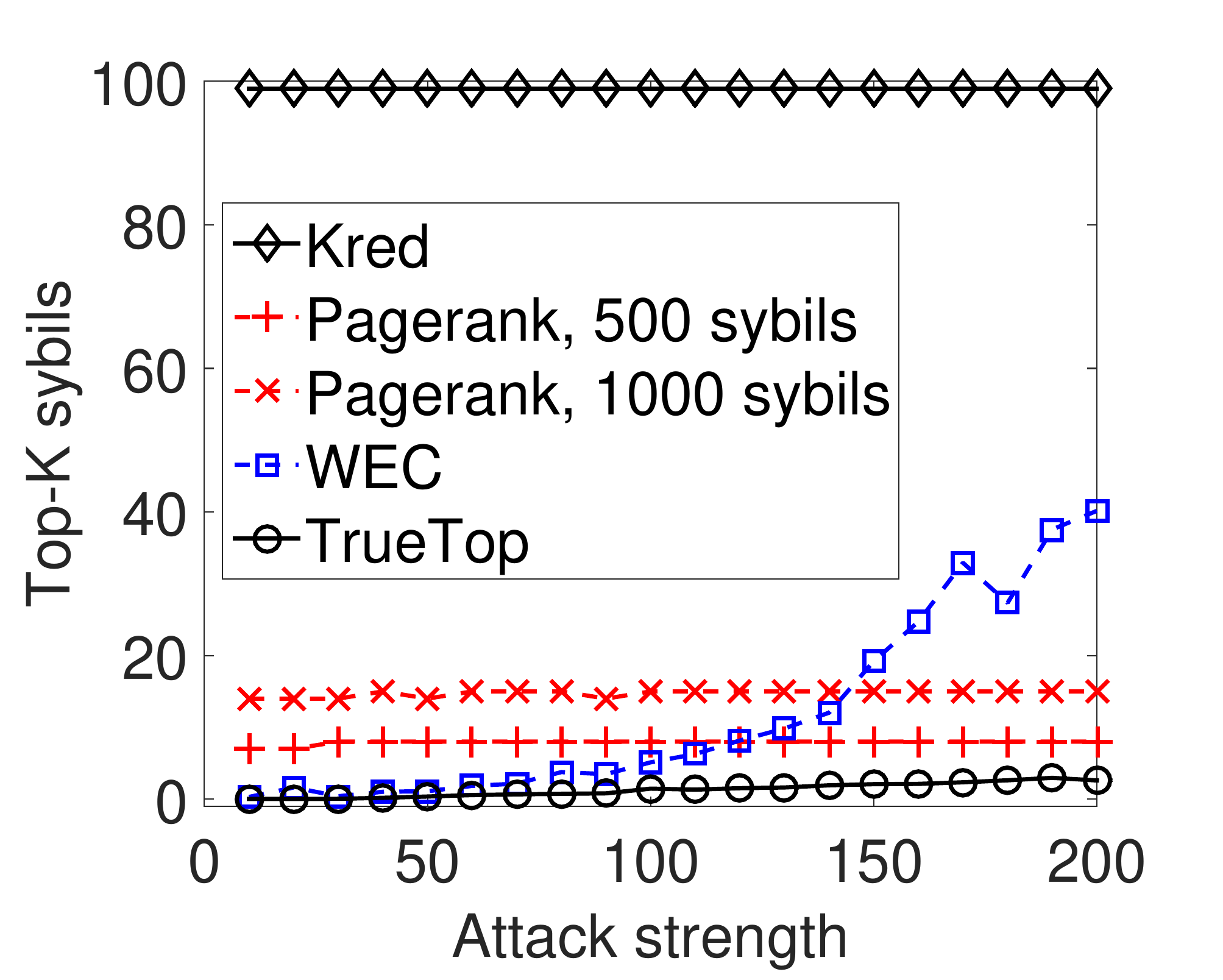}}\hfill
\caption{Comparing TrueTop with Kred, Pagerank and WEC with power iteration under the random and community attacks.}
\label{fig:sybil-compare}
\end{minipage}
\hspace{0.1cm}  
\begin{minipage}{0.24\textwidth}
\centering
\includegraphics[width=1.0 \textwidth]{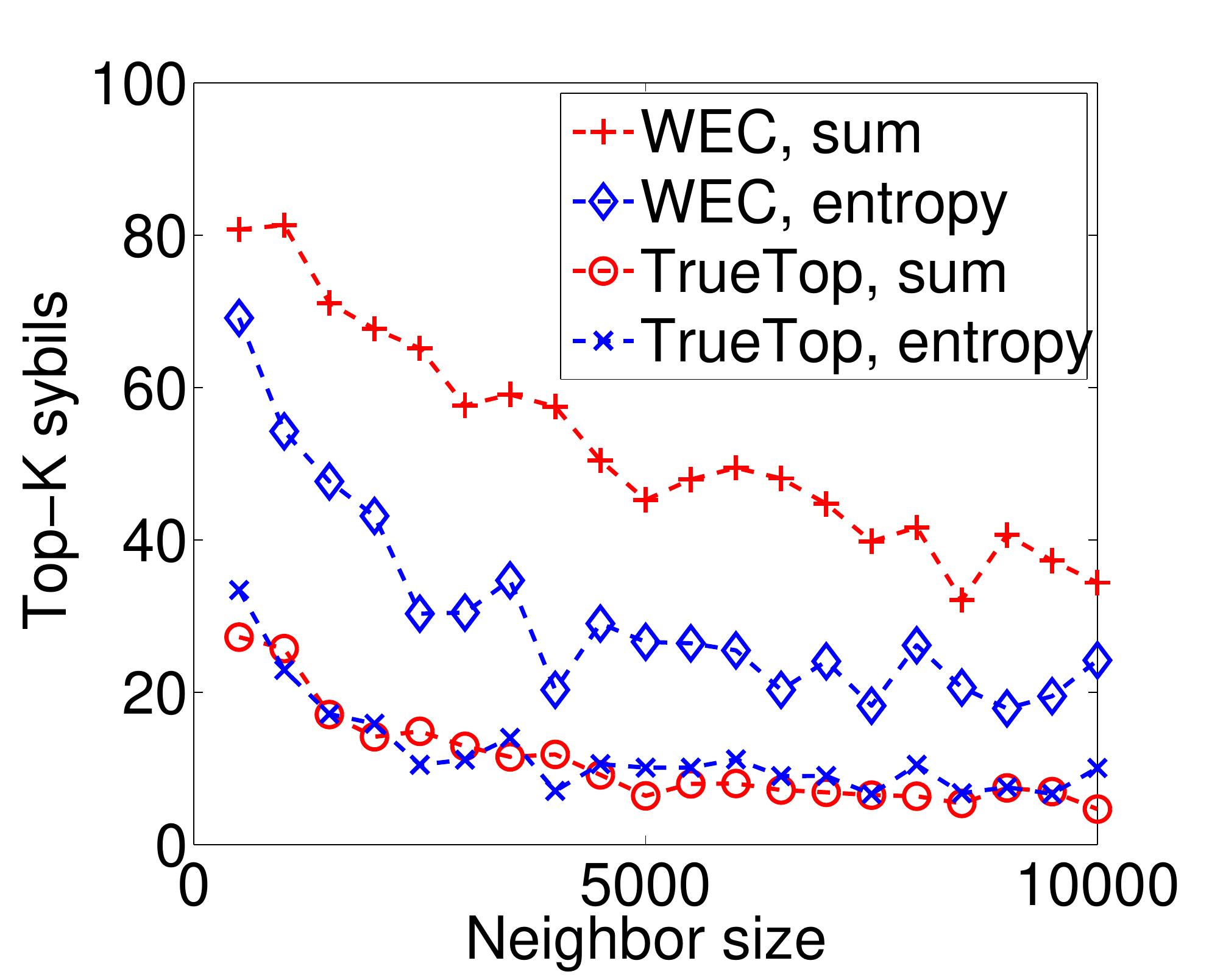}
\caption{TrueTop and WEC under seed attacks. }
\label{fig:neighbor}
\end{minipage}
\hspace{0.1cm}  
\begin{minipage}{0.24\textwidth}
\centering
\includegraphics[width=1.0 \textwidth]{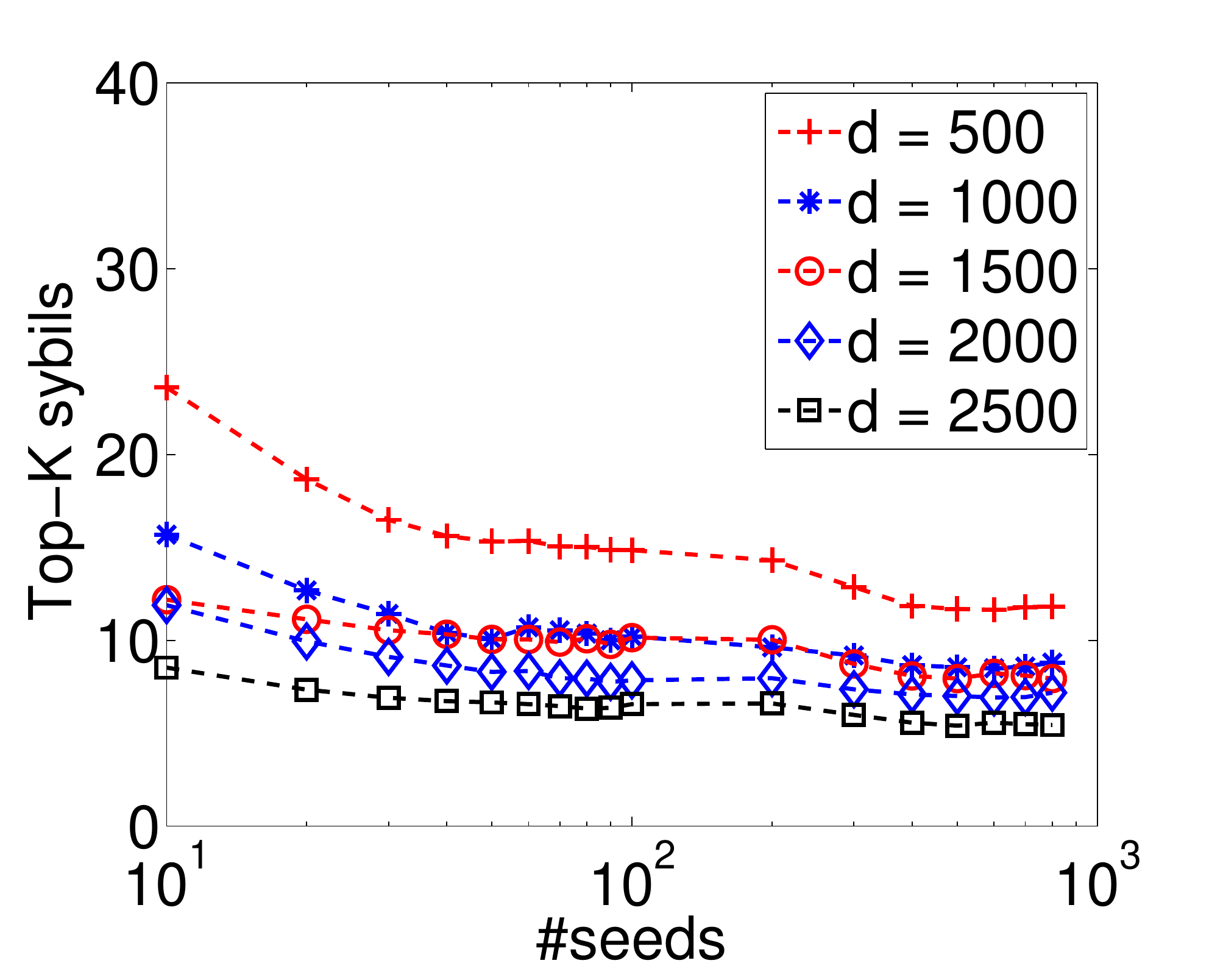}
\caption{Defense against the seed attack. }
\label{fig:neighborseeds}
\end{minipage}
\end{figure*}

Fig.~\ref{fig:sybil-weight} shows the performance of TrueTop under seed attacks for both sum-based and entropy-based edge weights. In this experiment, we set $K=100$ and $\epsilon=0$. In addition, we randomly selected $w_g$ users from $d=3,000$ immediate successors of 10 random seed users, from which $w_g$ links of weight one were added to the sybil region. We can have three observations from Fig.~\ref{fig:sybil-weight}. First, TrueTop is still very accurate as both type-I and type-II errors are always less than 2. Second, seed attacks can yield more sybil users in the top-$K$ list than both random and community attacks. Finally, entropy-based edge weights enable stronger sybil resilience than sum-based edge weights, as the former can dramatically increase the total edge weight in the non-sybil region in contrast to the total edge weight from the non-sybil region to the sybil region. An effective defense against the seed attack is deferred to Fig.~\ref{fig:neighborseeds}.

Table~\ref{tlb:option} shows the impact of design choices on the TrueTop performance. In this set of experiments, we set $K=100$, $\epsilon=0$, $w_g$ from 10 to 200, and $d=3,000$ for the seed attack. We compared the basic and reverse-WEC methods for seed selection, sum-based and entropy-based methods for determining edge weights, and also 10 versus 100 seed users. For simplicity, we added up the type-I errors, type-II errors, and $\#_{\textsf{sybil}}$ values under different attack strengths for each design choice, respectively. For each pair of design choices, we subtracted the sum of the second choice from that of the first one for the type-I error, type-II error, and $\#_{\textsf{sybil}}$, respectively. Since most results in Table~\ref{tlb:option} are positive, it is clear that the second choice in each pair can achieve higher accuracy and sybil resilience in most cases. Specifically, as expected, the entropy-based weight model yields better sybil resilience performance than the sum-based model.

\subsubsection{Comparison with Other Methods}

We compare our algorithm with the following methods.
\begin{enumerate}
\item \emph{Kred} \cite{Kred}. Since Kred has publish its influence score algorithm on \url{http://kred.com/rules}, we select it as the benchmark mechanism. Kred only computes the influence score by how many interactions a user have received in the past 1,000 days. During our 90-days experiment, we let each of the 500 sybils retweet each other sybil once per day. Therefore, each sybil receives 44,910 interactions from the sybils in the end. We will see that this conservative attack is sufficient for filling the top-$K$ list with mostly sybils.

\item \emph{Pagerank} \cite{PagePag99}. One may think about using the Pagerank value of each user in the interaction graph to evaluate his influence. Modified power iteration with non-zero reset probability is commonly used to compute Pagerank values. We set the rest probability to 0.15.
\item \emph{WEC by power iteration}. This method corresponds to TrueTop without early termination. 
\end{enumerate}

Fig.~\ref{fig:sybil-compare} compares the number of top-100 sybils of TrueTop with those of Kred, Pagerank and WEC by power iteration. As we can see, TrueTop allows less than 4 sybil users in the top-$100$ list under both random and community attacks. By comparison, the sybils in Kred can easily occupy 99 positions of the top-100 list. We also expect they will occupy all the top-100 positions if more interactions between the sybils were conducted. This is because the sybils can obtain unlimited incoming interactions from other sybils. Under WEC with power iteration, sybil users can occupy a significant portion in the top-100 list, as a lot more credits flow into and stay in the sybil region when power iteration terminates in contrast to TrueTop. In addition, Pagerank leads to more top-100 sybil users than TrueTop and is less sensitive to the attack strength than WEC with power iteration. However, if we increase the number of sybil users from 500 to 1,000 without changing the attack strength, the top-100 sybil users under Pagerank will increase. This is because the more sybil users, the higher probability that credit distribution jumps to the sybil region due to resetting operations, the higher Pagerank values of some sybil users. So Pagerank is not sybil-resilient either, which is consistent with \cite{ChengMan06}. In contrast, both TrueTop and WEC with power iteration are insensitive to the size of the sybil region.

Since WEC with power iteration is equivalent to seed-based iterative credit distribution without early termination, we also compare it with TrueTop with regard to the resilience to the seed attack. Note that Pagerank is not vulnerable to the seed attack because it does not use any seed user. Fig.~\ref{fig:neighbor} compares the top-100 sybil users of the two methods under the seed attack, where the number of immediate successors of the 10 victim seed users varies from $d=5000$ to $10,000$ for the fixed attack strength $w_g=100$. As we can see, both methods yield more top-100 sybil users as $d$ increases under sum-based and also entropy-based edge weights. This result is quite intuitive: the smaller $d$, the fewer nodes sharing the initial credits from the seed users, the more credits flowing into the sybil region over the $w_g$ links, and vice versa.

An effective defense again the seed attack is to select more seed users and/or choose the verified users with more immediate successors as seed users. The efficacy of this defense is shown in Fig.~\ref{fig:neighborseeds}. In this experiment, we assume that the attacker picked up 10 random seed users and then randomly selected $d$ immediate successors of them for adding the $w_g$ links to the sybil region. We varied the number of seeds from 10 to 800 for each value of $d$. As we can see, we can dramatically improve the resilience of TrueTop to the seed attack by increasing both the number of seed users and the number of immediate successors of the seed users.

\subsubsection{Remarks}

We have three remarks on the performance evaluation above. First, our evaluation results demonstrate the lower-bound performance of TrueTop. Specifically, we adopted a very strong attacker model by assuming that the attacker withholds all the credits flowing into the sybil region by having zero interactions to the non-sybil region. In practice, sybil users often try to initiate interactions with non-sybil users for other purposes such as spamming and phishing than merely aiming to gain high influence scores. Therefore, we can expect fewer credits to stay in the sybil region than under our attacker model such that TrueTop shall have higher accuracy and sybil resilience in more practical settings. Second, we admit that our evaluations are not complete given so many design choices for TrueTop as shown in Table~\ref{tlb:option} and many possible attack strategies. We have only shown some important results here as the examples and expect similar results for other design choices and attack strategies. Finally, we modelled the sybil behavior in accordance with prior work \cite{MessiYou13, ZhangOn13, FerraRis14}. There are more advanced sybil attacks such as astroturfing \cite{RatkiTru11} which could attract more legitimate interactions from non-sybil users. Unfortunately, there is no efficient way to simulate such advanced sybil attacks on a large scale. Instead, we use high attack strength $w_g$ to model them in the experiment. As expected, TrueTop performs worse for higher $w_g$ but still shows better performance in contrast to other methods. The performance of TrueTop will certainly degrade if the sybils could completely mimic the behavior of legitimate users, but manipulating the sybils to behave so intelligently will involve huge adversarial effort. TrueTop can thus significantly raise the bar for attacks on influence measurement.

\section{Conclusion} \label{sec:conclusion}
Influential users are vital to accelerate large-scale information dissemination and acquisition on Twitter. In this paper, we presented TrueTop, the first sybil-resilient system to measure the influence of Twitter users to the best of our knowledge. Our theoretical studies and also performance evaluations confirmed the high accuracy and sybil resilience of TrueTop.

\section*{Acknowledgement}
We truly appreciate the anonymous reviewers for their constructive comments. This work was partially supported by US Army Research Office through W911NF-15-1-0328. The work of Chi Zhang was partially supported by the Natural Science Foundation of China under Grants 61202140 and 61328208, by the Program for New Century Excellent Talents in University under Grant NCET-13-0548, by the Innovation Foundation of the Chinese Academy of Sciences under Grand CXJJ-14-S132, and by the Fundamental Research Funds for the Central Universities under Grand WK2101020006.

\bibliographystyle{IEEETran}
\bibliography{IEEEabrv,wins,winsPub,OnlineReputation}

\setcounter{section}{1}

\section*{Appendix}

\subsection{Proof of Proposition~1}\label{sec:sybilVul}

\begin{proof}
Let the total credits in $\mathcal{H}$ and $\mathcal{S}$ at $t$-th iteration be $C^{(t)}_{\mathcal{H}}$ and $C^{(t)}_{\mathcal{S}}$, respectively. According to the credit distribution defined in Eq.~\ref{eq:credit}, after the $t$-th iteration, the average credits flowed from $\mathcal{H}$ to $\mathcal{S}$ and $\mathcal{S}$ to $\mathcal{H}$ are $\alpha C^{(t)}_{\mathcal{H}}$ and $\beta C^{(t)}_{\mathcal{S}}$, respectively. Meanwhile, the total credits in the whole network is constant to 1. Hence,
\begin{align*}
C^{(t)}_{\mathcal{H}} &= (1-\alpha)C^{(t-1)}_{\mathcal{H}} + \beta C^{(t-1)}_{\mathcal{S}}\\
                        &= (1-\alpha)C^{(t-1)}_{\mathcal{H}} + \beta (1 - C^{(t-1)}_{\mathcal{H}})\\
                        &= (1-\alpha - \beta)C^{(t-1)}_{\mathcal{H}} + \beta\\
                        &= (1-\alpha-\beta)^{t-1}C^{(1)}_{\mathcal{H}} + ((1-\alpha-\beta)^{t-2} + \ldots + 1)\beta\\
                        &= (\frac{1}{\alpha+\beta} - 1)\alpha(1-\alpha-\beta)^{t-1} + \frac{\beta}{\alpha+\beta}
\end{align*}
and
\[
C^{(t)}_{\mathcal{S}} = 1- C^{(t)}_{\mathcal{H}} = (1 - \frac{1}{\alpha+\beta})\alpha(1-\alpha-\beta)^{t-1} + \frac{\alpha}{\alpha+\beta}
\]
Since $\alpha \ll 1$ and $\beta \ll 1$, $C^{(t)}_{\mathcal{H}}$ will decrease monotonically and $C^{(t)}_{\mathcal{S}}$ will increase monotonically. When $t\rightarrow\infty$, $C^{(t)}_{\mathcal{S}} = \frac{\alpha}{\alpha+\beta}$.

\end{proof}

\subsection{Proof of Lemma~1}\label{sec:Lemma1}


\begin{proof}
According to the Perron-Frobenius theory \cite{BehreIntro02}, the matrix $\bf W$ is irreducible and has the largest eigenvalue of 1, and all other eigenvalues are absolutely less than 1, denoted as $1 = \lambda_1 > \lambda_2 \geq \lambda_3 \geq \ldots \geq \lambda_n > -1$. Moreover, if we denote the corresponding $n$ eigenvectors as ${\bf v}_1, {\bf v}_2, \ldots, {\bf v}_n$, then $|{\bf v}_1| = 1$ and we denote ${\bf v}_1$ as the WEC vector ${\boldsymbol\pi}$. Next if ${\bf W}$ is diagonalizable, then ${\bf v}_1, {\bf v}_2, \ldots, {\bf v}_n$ can be orthogonal to expand the whole space of $\mathbb{R}^n$. For the case of non-diagonalizable ${\bf W}$, we can use the Jordan canonical form to transform it into a diagonalizable one \cite{SchmiImp01}.

Since ${\bf v}_1, {\bf v}_2, {\bf v}_3, \ldots, {\bf v}_n$ are orthogonal, ${\bf v}_0$ can be written as
\begin{equation}
{\bf v}_0 = \sum_{i=1}^n a_i {\bf v}_i \label{eq:linear}
\end{equation}
where $a_i \in \mathbb{R}$. We argue that if ${\bf W}$ is stochastic and irreducible then $a_1 = 1$. To see why, we first notice that since ${\bf W}$ is stochastic, ${\bf W} \mathbf{1} = \mathbf{1}$. It follows that ${\bf v}^T_i {\bf W} \mathbf{1} = \lambda_i {\bf v}^T_i \mathbf{1} = {\bf v}^T_i \mathbf{1}$. The eigenvector corresponding to $\lambda_1$ is the stationary distribution of Markov Chain ${\bf W}$. Since ${\bf W}$ is irreducible, $\lambda_i < 1$ and $\lambda_i \neq 1$ when $i \neq 1$. Thus we can see that ${\bf v}^T_i \mathbf{1} = 0$ for $i \neq 1$. Multiplying $\mathbf{1}$ at both sides of Eq. \ref{eq:linear}, it follows that ${\bf v}_0 \mathbf{1} = a_1 {\bf v}_1 \mathbf{1}$. Since both ${\bf v}_0$ and ${\bf v}_1$ are non-negative vectors with the sum of 1, we have $a_1 = 1$.

Thus Eq. \ref{eq:linear} can be simplified as
\[
{\bf v}_0 = {\bf v}_1 + \sum_{i=2}^n a_i {\bf v}_i = {\boldsymbol\pi} + \sum_{i=2}^n a_i {\bf v}_i\;.
\]

Multiplying ${\bf W}^t$ at both sides and keeping using the equation ${\bf v}_i {\bf W} = \lambda_i {\bf W}$, we can obtain
\[{\bf x}^{(t)} =  {\bf v}_0 {\bf W}^t = ({\boldsymbol\pi} + \sum_{i=2}^n a_i {\bf v}_i){\bf W}^t = {\boldsymbol\pi} + \sum_{i=2}^n \lambda^t_i a_i {\bf v}_i\;.\]

Let $\lambda = max(|\lambda_2|, |\lambda_n|)$. As the $t\rightarrow \infty$, $\lambda^t$ will become dominant and it follows that $|({\bf x}^{(t)} -{\boldsymbol \pi})_i| = O(\lambda^t)$

Moreover, for $j\in \mathcal{U} \setminus {\bf s}, \sum_{i=1}^n a_i v_{i,j} = v_{0,j} = 0$. Hence,
\[
e^{(t)}_j  = |\sum_{i=2}^n \lambda^t_i a_i v_{i,j}|  \leq \lambda^t|\sum_{i=2}^n a_i v_{i,j}| =\lambda^t|- v_{1,j}| =\lambda^t \pi_j\;. \]
\end{proof}

\subsection{Proof of Theorem~1}\label{sec:Theorem1}

\begin{proof}
The conclusion is composed of two parts. We begin with the first part, i.e., if $\lambda^t \leq \Delta'_k/2$ at the $t$-th iteration, then $x_1 > x_2 > \ldots \>x_k$. Consider the $k$- and $(k-1)$-ranked nodes. Since the relative WEC gap $\Delta'_k$ is monotone decreasing for $k$, we have
\[
e'_k \leq \lambda^t \leq \Delta'_k / 2 < \Delta'_{k-1} / 2
\]
Combined with $e'_{k-1} \leq \lambda^t$ in Lemma \ref{th:error}, we can get $e'_{k - 1} < \Delta'_{k-1} / 2$. In other words,
\[
\left\{
\begin{array}{l}
e_k = |x_k -\pi_k| \leq \Delta_k / 2, \\
e_{k-1} = |x_{k-1} -\pi_{k-1}| < \Delta_{k-1} / 2.
\end{array}
\right.
\]
By several operations, we have
\[
x_{k-1} -x_k > ((\pi_{k-1} - \pi_k) - (\pi_k - \pi_{k+1} )) / 2 > 0
\]
which holds since $\Delta_k / \pi_k < \Delta_{k-1} / \pi_{k-1} < \Delta_{k-1} / \pi_k$. Similarly, we can find that $x_{k-2} > x_k, \ldots, x_1 > x_k$. Moreover, if starting from $(k-1)$-ranked node (it holds as $e'_{k - 1} \leq \lambda^t \leq \Delta'_{k-1} / 2$), we have $x_{k-2} > x_{k-1}, \ldots, x_1 > x_{k-1}$ and thus $x_1 > x_2 > \ldots > x_k$.

Then we prove the second part, i.e., if $\lambda^t \leq \Delta'_k/2$ at the $t$-th iteration, then $x_k > x_{j}$, where $j$ is from $k+1$ to $n$. Consider the $k$- and $(k+1)$-ranked nodes. We have
\[
\left\{
\begin{array}{l}
e_{k+1} \leq \lambda^t \pi_{k+1} <  \lambda^t \pi_{k} \leq \Delta_{k}  / 2,\\
e_k \leq \lambda^t \pi_{k} \leq \Delta_{k}/ 2.
\end{array}
\right.
\]
Hence,
\[
x_{k+1} - x_k < 0
\]
and so for all other nodes with the rankings larger than $k$.

Since $\lambda^t$ is geometrically decreasing for $t$, $\lambda^t  < \Delta'_k/2$ holds for all the following iterations and so does the conclusion.
\end{proof}

\subsection{Proof of Theorem~2}\label{sec:Theorem2}
\begin{proof}
According to \cite{GhoshRan11}, the expected $k$-ranked WEC is
\[
\langle \pi \rangle_k \approx \frac{\Gamma(k-\frac{1}{\gamma - 1})}{\Gamma(k)}
\]
According to Proposition~\ref{th:credit}, the number of credits for $\mathcal{S}$ after the $t$-th iteration is given by:
\[
C^{(t)}_{\mathcal{S}} = 1- C^{(t)}_{\mathcal{H}} = (1 - \frac{1}{\alpha+\beta})\alpha(1-\alpha-\beta)^{t-1} + \frac{\alpha}{\alpha+\beta}
\]
The maximum $C^{(t)}_{\mathcal{S}}$ can be obtained as $1 - (1-\alpha)^t$ when $\beta \rightarrow 0$, i.e., the sybils conduct very limited interactions to the non-sybil users. Moreover, since the attacker wants to place as many sybils into the top-$K$ list as possible, he can just divide the total credits $C^{(t)}_{\mathcal{S}}$ by the $k$-ranked WEC value. Then the number of sybils that own $\langle \pi \rangle_K$ credits is given by
\[
n(K) = \frac{ C^{(t)}_{\mathcal{S}}}{C^{(t)}_{\mathcal{H}} \langle \pi \rangle_K} \approx \frac{1 - (1-\alpha)^t}{(1-\alpha)^t\langle \pi\rangle_K}
\]

Here we further approximate the $\langle \pi \rangle_K$. For the power law distribution, $2\leq\gamma < 3$. Thus
\[
\frac{\Gamma(k-\frac{1}{\gamma - 1})}{\Gamma(k)} > \frac{\Gamma(k-1)}{\Gamma(k)} =  \frac{1}{k}
\]
Hence, we can obtain $n(K) < K(1 - (1-\alpha)^t) / (1-\alpha)^t$.
\end{proof}

\begin{IEEEbiography}[{\includegraphics[width=1in,height=1.25in,clip,keepaspectratio]{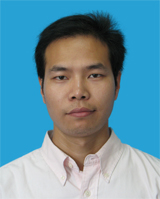}}]{Jinxue Zhang}
is currently a Ph.D. student in Electrical, Computer and Energy Engineering at Arizona State University. He received the M.E. in Software
Engineering from Tsinghua University at 2011 and B.E. in Communication Engineering from Nanjing University at 2006, both from China.
His research focuses on security and privacy issues in online social networks.
\end{IEEEbiography}

\begin{IEEEbiography}[{\includegraphics[width=1in,height=1.25in,clip,keepaspectratio]{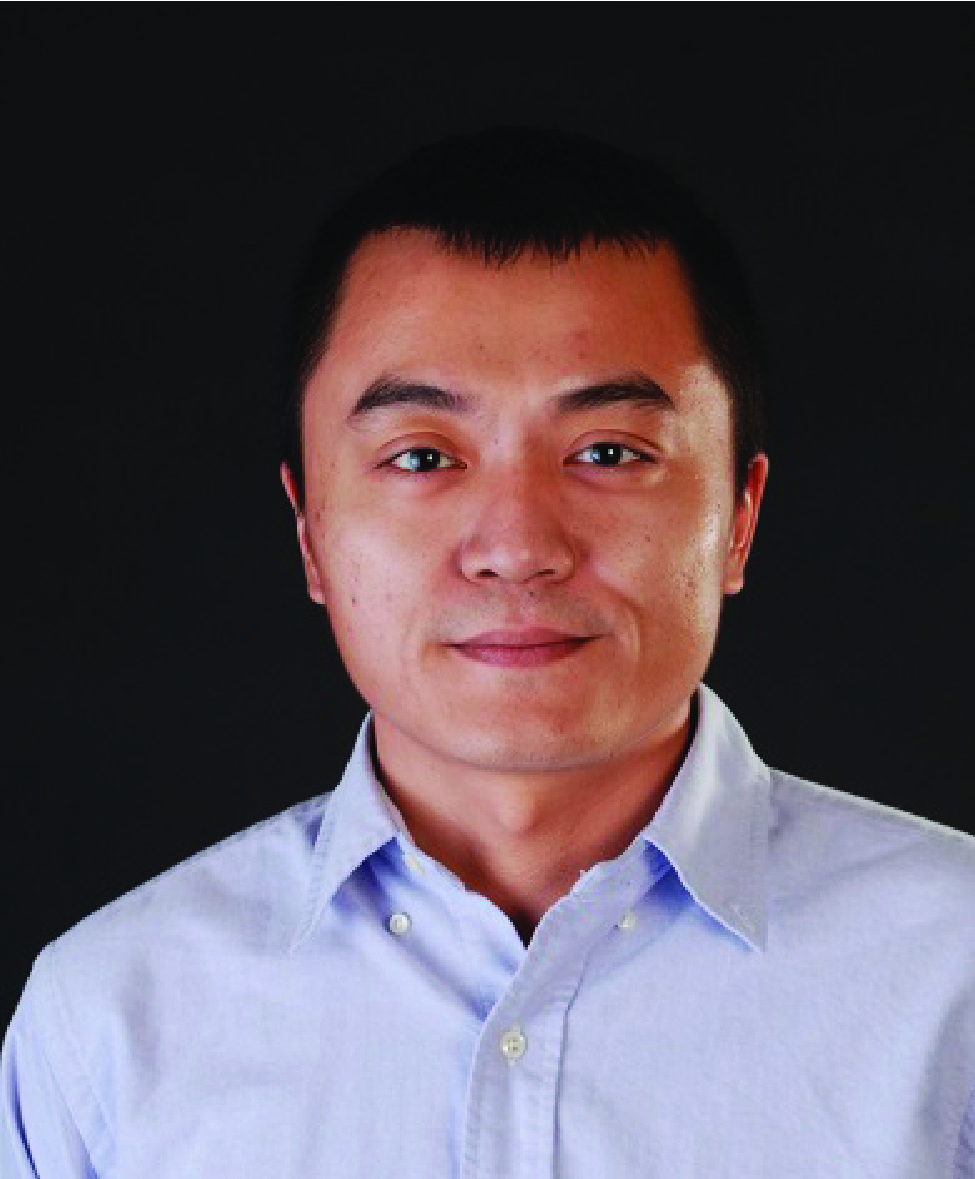}}]{Rui Zhang}
received the B.E. in Communication Engineering and the M.E. in Communication and Information System from Huazhong University of Science and
Technology, China, in 2001 and 2005, respectively, and the PhD degree in electrical engineering from the Arizona State University, in 2013.
He was a software engineer in UTStarcom Shenzhen R$\&$D center from 2005 to 2007. He has been an assistant professor in the Department of
Electrical Engineering at the University of Hawaii since July 2013. His primary research interests are network and distributed system
security, wireless networking, and mobile computing.
\end{IEEEbiography}

\begin{IEEEbiography}[{\includegraphics[width=1in,height=1.25in,clip,keepaspectratio]{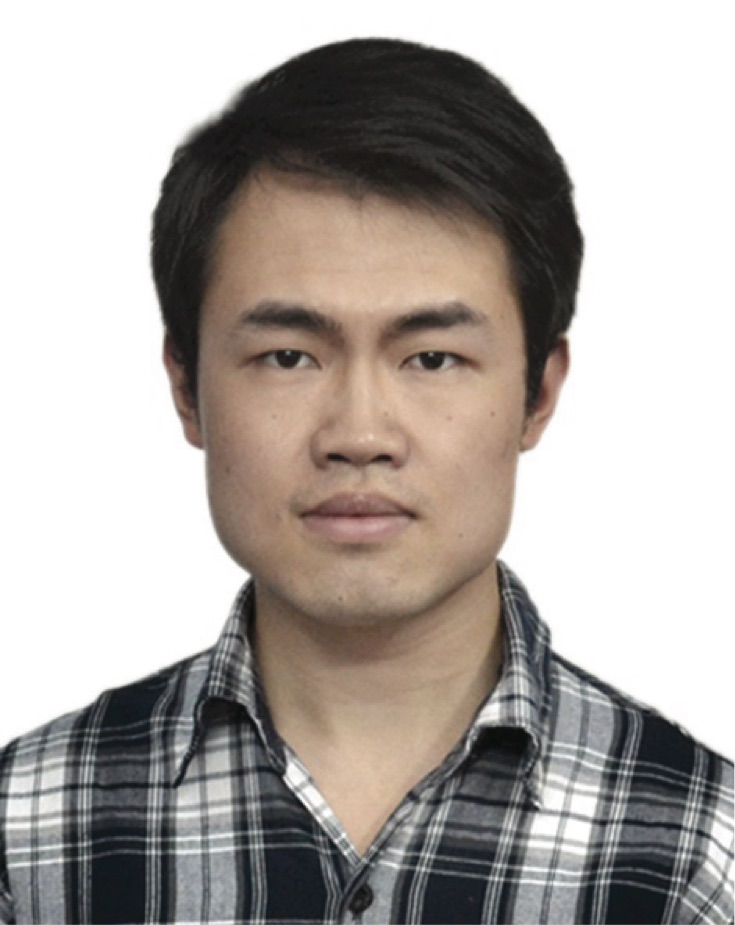}}]{Jingchao Sun}
received the B.E. in Electronics and Information Engineering and the M.E. in Communication and Information System from Huazhong
University of Science and Technology, China, in 2008 and 2011, respectively. He is currently a Ph.D. student in School of Electrical,
Computer, and Energy Engineering at Arizona State University. His primary research interests are network and distributed system
security and privacy, wireless networking, and mobile computing.
\end{IEEEbiography}

\begin{IEEEbiography}[{\includegraphics[width=1in,height=1.25in,clip,keepaspectratio]{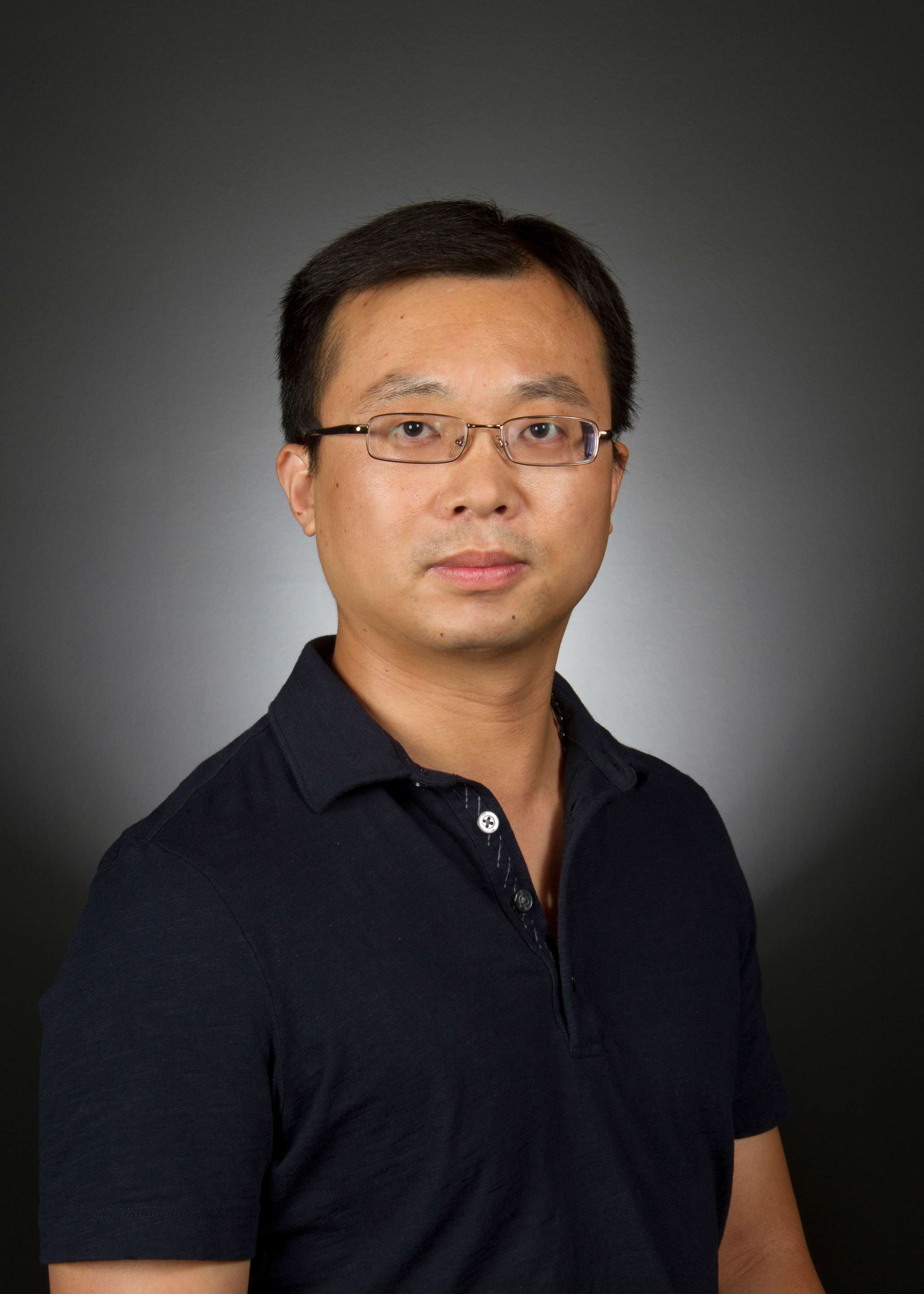}}]{Yanchao Zhang}
received the B.E. in Computer Science \& Technology from Nanjing University of Posts \& Telecommunications in 1999, the M.E. in Computer Science \& Technology from Beijing University of Posts \& Telecommunications in 2002, and the Ph.D. in Electrical and Computer Engineering from University of Florida in 2006. He came to Arizona State University in June 2010 as an Associate Professor of School of Electrical, Computer and Energy Engineering and the director of ASU Cyber \& Network Security Group (CNSG).  Prior to joining ASU, he was an Assistant Professor of Electrical and Computer Engineering at New Jersey Institute of Technology from August 2006 to June 2010. His primary research is about security and privacy issues in computer and networked systems, with current focus areas in emerging wireless networks, mobile crowdsourcing, Internet-of-Things, social networks, wireless/mobile systems for disabled people, mobile and wearable devices, and wireless/mobile health. He is/was on the editor boards of IEEE Transactions on Mobile Computing, IEEE Wireless Communications, IEEE Transactions on Control of Network Systems, and IEEE Transactions on Vehicular Technology. He received the US National Science Foundation Faculty Early Career Development (CAREER) Award in 2009.
\end{IEEEbiography}

\begin{IEEEbiography}[{\includegraphics[width=1in,height=1.25in,clip,keepaspectratio]{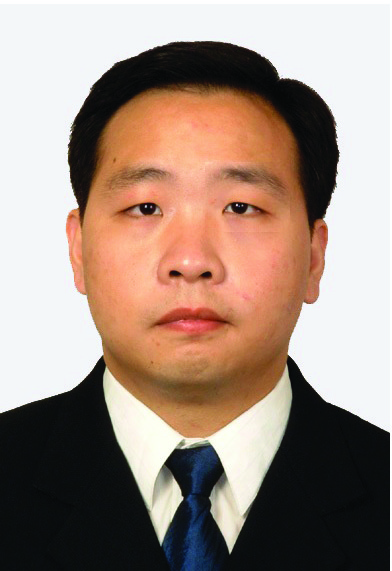}}]{Chi Zhang}
received the B.E. and M.E. in electrical and information engineering from Huazhong University of Science and Technology, Wuhan, China, in 1999 and 2002, respectively, and the Ph.D. in electrical and computer engineering from the University of Florida, Gainesville, Florida, in 2011. He joined University of Science and Technology of China in September 2011 as an Associate Professor of School of Information Science and Technology. His research interests are in the areas of network protocol design, network performance analysis, and network security guarantee, particularly for wireless networks and social networks. He received the 7th IEEE ComSoc Asia-Pacific Outstanding Young Researcher Award.
\end{IEEEbiography}

\end{document}